\documentclass[a4paper,11pt]{article}

\usepackage{jinstpub} 
 \pdfoutput=1
\usepackage{lscape}
\usepackage{lineno}
\usepackage{subcaption}
\usepackage{tablefootnote}
 \usepackage{hyperref} 
 \usepackage{url}
\usepackage{booktabs, caption, makecell}
\usepackage{threeparttable}

\title{\boldmath Positron sources: from conventional to advanced  accelerator concepts-based colliders
}

\author[a,1]{I. Chaikovska,\note{Corresponding author.}}
\author[a]{R.~Chehab,}
\author[a]{V.~Kubytskyi,}
\author[a]{S.~Ogur,}
\author[a]{A.~Ushakov,}
\author[b]{A.~Variola,}
\author[c]{P.~Sievers,}
\author[d]{P. Musumeci}
\author[e]{L.~Bandiera}
\author[f]{Y.~Enomoto,}
\author[g]{Mark~J.~Hogan}
\author[h]{P. Martyshkin}

\affiliation[a]{Université Paris-Saclay, CNRS/IN2P3, IJCLab, 91405 Orsay, France}
\affiliation[b]{INFN Rome Unit, P.le A. Moro 1, 00185 Rome, Italy}
\affiliation[c]{European Organization for Nuclear Research (CERN), Geneva, Switzerland}
\affiliation[d]{Department of Physics and Astronomy, University of California at Los Angeles}
\affiliation[e]{INFN Ferrara Unit, via Saragat 1, 44122 Ferrara, Italy }
\affiliation[f]{High Energy Accelerator Research Organization (KEK), Oho 1-1, Tsukuba, Ibaraki, 305-0801, Japan}
\affiliation[g]{SLAC National Accelerator Laboratory, 2575 Sand Hill Road, Menlo Park, California 94025, USA}
\affiliation[h]{BINP SB RAS, Novosibirsk, Russia}

\emailAdd{iryna.chaikovska@ijclab.in2p3.fr}

\abstract{ 
Positron sources are the key elements for the future and current lepton collider projects such as ILC, CLIC, SuperKEKB, FCC-ee, Muon Collider/LEMMA, etc., introducing challenging critical requirements for high intensity and low emittance beams in order to achieve high luminosity. In fact, due to their large production emittance and constraints given by the target thermal load, the main collider parameters such as the peak and average current, the emittances, the damping time, the repetition frequency and consequently the luminosity are determined by the positron beam characteristics.
In this paper, the conventional positron sources and their main properties are explored for giving an indication to the challenges that apply during the design of the advanced accelerator concepts.
The photon-driven positron sources as the novel approach proposed, primarily for the  future linear colliders, are described highlighting their variety and problematic.}

\keywords{Accelerator Subsystems and Technologies; Targets; Positron Sources; Accelerator Applications}

\begin{document}
\maketitle
\flushbottom


\section{Introduction}
\label{sec:intro}
In view of the update of European (EU) High-Energy Physics (HEP) strategy for particle physics, the highest scientific priorities are given by the precision study of the Higgs boson and the exploration of the high-energy frontier~\cite{europeanstrategygroup2020physics, adolphsen2022european}. These are two complementary ways to address the open questions in particle physics. In this context, a clear indication has been given to explore the feasibility of an electron-positron ($e^-e^+$) collider, since lepton colliders should provide high precision measurements. Nowadays, there are four possible projects: the International Linear Collider (ILC) in Japan, the Compact Linear Collider (CLIC) or the Future Circular Collider (FCC-ee) at CERN  and the Circular Electron Positron Collider (CEPC) in China. 
In this context, of great importance are also the ongoing feasibility studies on the future multi-TeV $\mu^+\mu^-$ Muon Collider, which can serve for both precision measurements and discovery physics \cite{Bartosik_2020}.
These are the future pioneering accelerators of the post-LHC era and the related physics program is extremely ambitious \cite{Peach, FCC:2018byv,  clic_phys}.

In this framework, an important role should be played by the field of advanced accelerators, including laser and beam driven plasma and advanced structure concepts \cite{adolphsen2022european}, especially to reduce dimensions and costs of the electron primary beams and of the positrons post acceleration sections. In the last decade the field has seen tremendous progress with the demonstration of multi-GeV acceleration in a single stage, the first staging of plasma accelerators, and greatly improved beam quality. At the same time, there are still many open questions to be addressed before compact linear colliders based on advanced accelerators could be built, first of all to provide the necessary beam intensity to fulfill the luminosity requirements. With this goal in mind, two particular directions can be identified. On one side, research on advanced accelerator techniques should be pursued to solve the outstanding critical issues in order to enable the next generation high-energy physics machine. At this regard, the development of an integrated design study for compact high-gradient colliders is critical to guide the efforts and provide a clear and actionable R\&D path. On the other side, it is also clear, at present, the need to identify and pursue nearer-term applications both inside and outside high-energy physics. Successful deployment of advanced techniques in real-world accelerator applications will be essential to strengthen the case for this field and provide the necessary intermediate steps before deployment of compact accelerators for the most demanding high-energy physics applications.  



\section{Positron sources for High-Energy Physics applications}
\label{sec:HEP_applications}

Nowadays, positron beams and associated positron sources have extremely versatile applications, ranging from high energy physics to atomic, solid state physics and their use in a nuclear medicine imaging procedure called PET (Positron Emission Tomography). 
While their beam parameters and design vary from application to application, we can state that a current increasing demand is evident for beams with higher intensity and lower emittances.

The positrons are usually produced in $\beta^+$ decays of radioactive isotopes or in pair conversion by energetic photons on the nuclear potential of a target. 
The positrons originated from the $\beta^+$ decay are polarized, have a broad energy distribution and emitted isotropically with a random time structure, which make them difficult to be captured and accelerated. 
In addition, the intensity provided by the radioactive sources is $\sim10^6-10^8$ $e^+$/s~\cite{BAUER1990300, golge2012review}, which is 4 to 6 orders of magnitude lower than usually required by the high-energy physics applications. In fact, cold positron plasma with good phase space characteristics can be produced by accumulating positrons emitted by radioactive source in special devices~\cite{cassidy2006accumulator}, but the long accumulation time implicitly forbids the extension of their possible use for the HEP application. 
In this context, for the GBAR (Gravitational Behaviour of Antihydrogen at Rest) experiment at CERN a source of slow positrons based on a low-energy (9~MeV) electron linear accelerator has been constructed. Positrons are first trapped
in a buffer-gas accumulator and then collected in a high-field (5~T) Penning–Malmberg trap. In order to be trapped the positron energy must be in the eV–keV range. The system provides $5\times 10^7$ $e^+$/s positron flux, which is fed into a buffer-gas trap~\cite{CHARLTON2021164657}.
Conversion of the trapped positron plasma by using an electrostatic traps into a usable beam, which can be injected into many kinds of accelerator has been recently explored~\cite{hessami:napac2019-moplh23}. 
In such a way, the interaction of relativistic particles or photon beams with matter is used when designing the HEP positron sources at large-scale facilities. Thus, the positron beams are always secondary beams.

In the framework of the design and realization of a lepton collider, positron sources are essential due to the challenging critical requirements of high-beam intensity and low emittance necessary to achieve high luminosity. 
In this scheme, being produced in the target, the generated positron beams have a very broad angular and energy distributions. Thus, owing to their large production emittance and intensity constraints given by the target thermal load, the main collider parameters like the peak and average current, the final emittances, the damping time, the repetition frequency and consequently the luminosity are determined by the positron beam characteristics.

Another option of positron sources or positron accelerators can incorporate or use in their design novel concepts, which employ the wakefields excited in a plasma or accelerating structures (which can be made of dielectric material) for acceleration of charged particles. 
It includes two beam-driven techniques, where a beam passes through a plasma
or a structure: Plasma WakeField Acceleration (PWFA) and Structure-based WakeField Acceleration (SWFA) respectively; and two externally powered acceleration methods, where an external electromagnetic pulse (optical laser or THz pulse) is coupled to a plasma or structure to accelerate the beam: Laser WakeField Acceleration (LWFA) and Dielectric Laser Acceleration (DLA) respectively.
In the last decades, it was demonstrated that plasma-based accelerators have made rapid technical and conceptual progress and their performance is getting closer to those suitable for future colliders. 
However, the feasibility of a collider based on plasma accelerator schemes still remains to be proven.
Although such novel techniques are very promising for electrons, a scheme for positron bunch acceleration in plasma still needs to be demonstrated for applications towards future high-energy colliders. 

Positron beam experiments performed at the SLAC's Final Focus Test Beam (FFTB) demonstrated the first positron acceleration in plasma~\cite{PhysRevLett.90.214801}. It showed that the positron beam was distorted after passing through a low density plasma. Several experimental activities were undertaken at Facility for Advanced Accelerator Experimental Tests (FACET) on the acceleration of injected positron bunches in a beam-driven plasma accelerator ~\cite{PhysRevLett.90.205002,PhysRevLett.101.055001}.
A few positron acceleration concepts/techniques have been, thus, emerged: using the acceleration in quasi-linear or non-linear wakefields~\cite{doche2017acceleration, corde2015multi} and hollow channel plasma wakefield acceleration~\cite{gessner2016demonstration}.
An overview and future plans to improve the efficiency and beam quality for positron acceleration has been reported~\cite{PhysRevResearch.3.043063}.
In this context, the proposed upgrade of FACET (FACET-II) is the only facility that can provide the high-energy positron beams for the experimental study of positron acceleration in plasma and, potentially, multi-stage positron acceleration. Recently, the implication of the laser-driven plasma accelerators to positron sources has been reviewed showing the possibility of using the laser-driven positron beams as a witness beam for beam-driven and laser-driven wakefield test facilities~\cite{alejo2019laser}.


\subsection{Conventional approach and state of the art}
\label{sec:state_of_art}
\begin{figure}[h]
\centering
\begin{subfigure}{.5\textwidth}
  \centering
  \includegraphics[width=1.\linewidth]{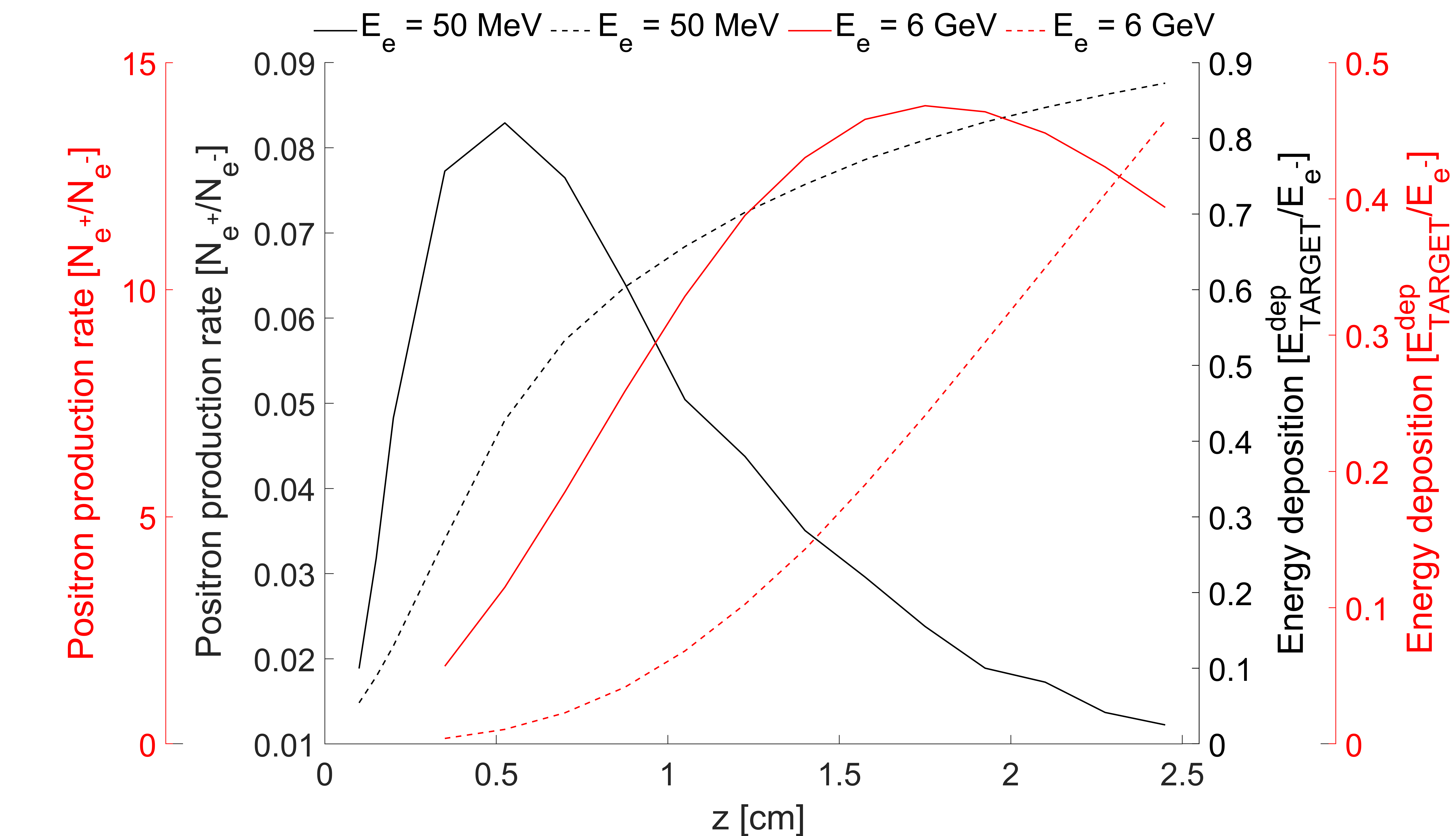}
  \caption{}
  \label{fig:sub1}
\end{subfigure}%
\begin{subfigure}{.5\textwidth}
  \centering
  \includegraphics[width=1.\linewidth]{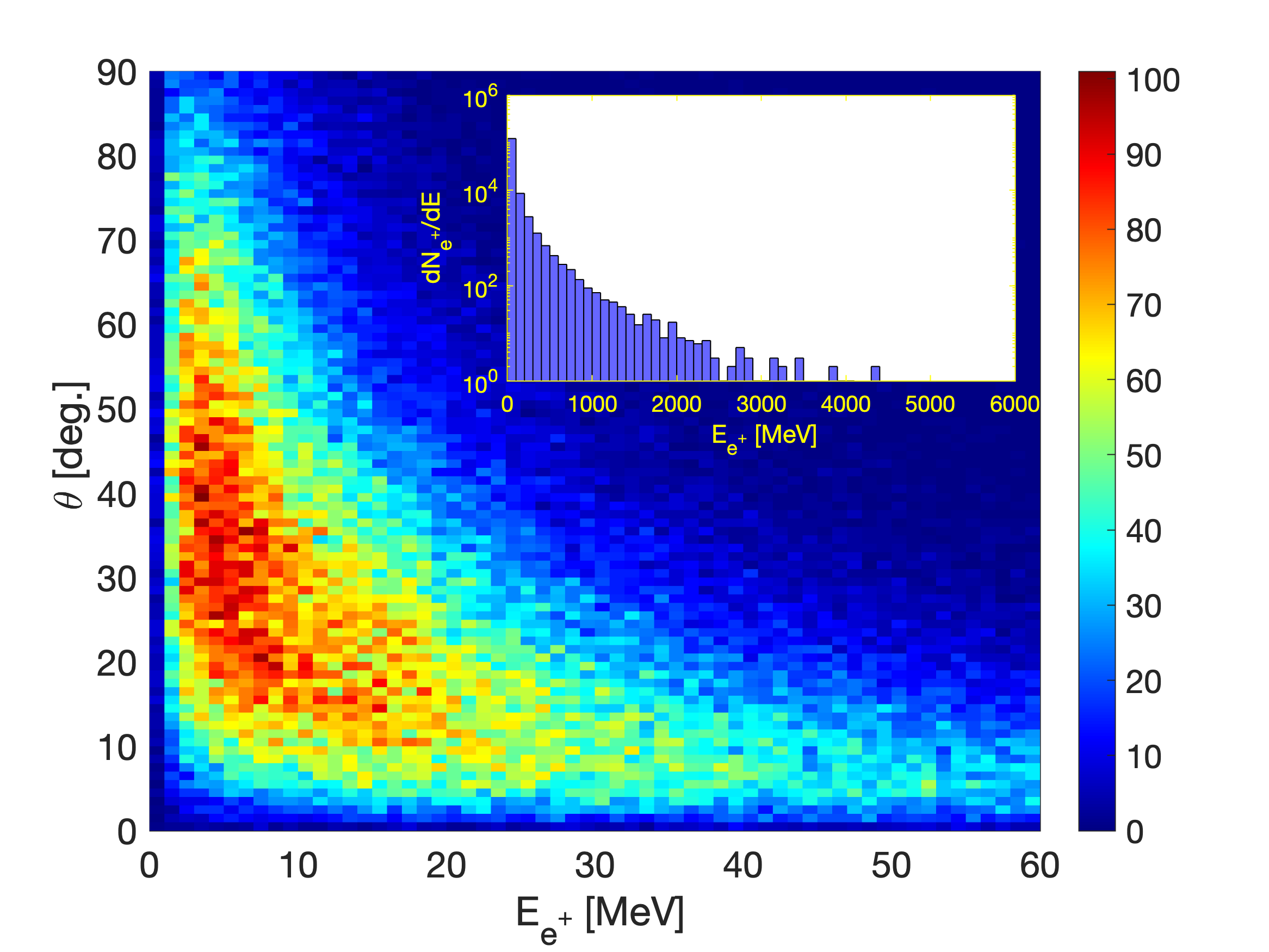}
  \caption{}
  \label{fig:sub2}
\end{subfigure}
\caption{Characteristics of the positrons produced in the conventional scheme. (a) Positron production rate and fraction of energy deposited in the target as a function of target thickness for 50~MeV and 6~GeV electron drive beam energy. (b) Angular-energy distribution of the positrons at the exit of the target-converter for the 6~GeV electron drive beam. Note, that in both cases the tungsten (Z=74) target-converter has been used in simulations.}
\label{fig:Positrons_dist}
\end{figure}

 In all positron sources used for accelerators, positrons are produced as secondary beams from the electromagnetic shower cascade generated by high-energy electrons hitting a target of high-Z material allowing relatively high conversion efficiency. 
 In this process the particle/matter interactions dominate fixing the source parameters. The resulting final 6D~normalized emittance is orders of magnitude higher than in high-brilliance electron sources due to the large momentum and angular spread generated by the shower processes and the multiple scattering in the target. The interplay between the drive beam energy, target thickness and positron production rate as well as the typical angular/energy distributions of the produced positrons are illustrated in Figure~\ref{fig:Positrons_dist}. 
 In general, there are two main aspects that affects the positron production rate: incident energy/intensity and target characteristics (material, thickness).
 
\begin{figure}[h]
\centering
\includegraphics[width=0.5\textwidth]{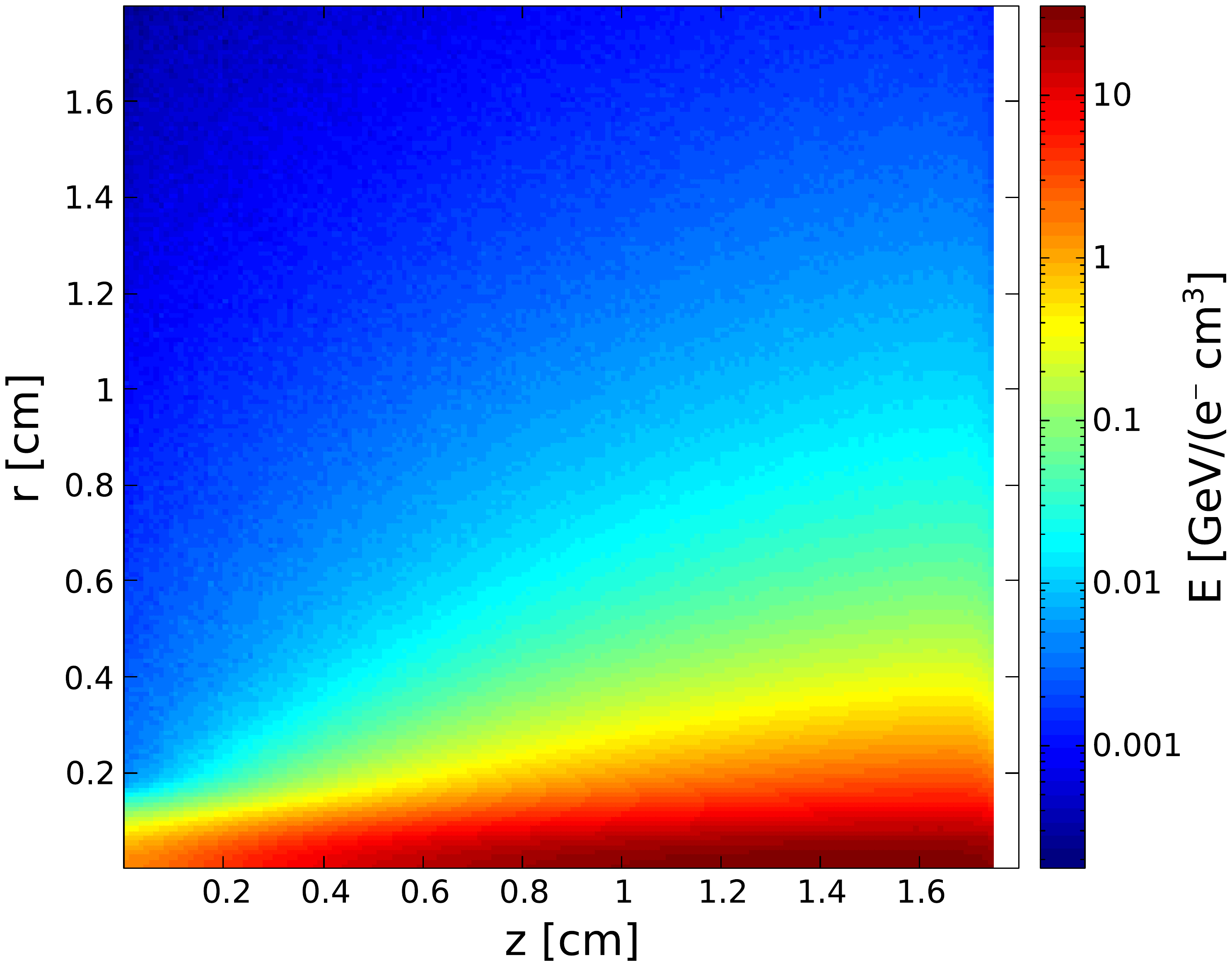}
\caption{ Energy deposition map in the tungsten target-converter. The maximum value (PEDD) is reached at the exit of the target, where the electromagnetic shower is at its maximum. Simulation was performed for the electron beam energy of 6~GeV and a beam spot size on the target being~0.5~mm.}
\label{fig:positronsPEDD}
\end{figure}

In the production process, the primary beam energy deposition density in the converter is far from being homogeneous (see Figure~\ref{fig:positronsPEDD}), leading to a high thermo-mechanical stress, given by the temperature gradient produced in the target, and a related target failure threshold. Therefore, an appropriate parameter, so-called, the Peak Energy Deposition Density (PEDD) has been introduced in positron source design to provide a quantitative parameter assessing the reliability of the target operation \cite{Bharadwaj2001slc}.

After the pair generation, due to their large transverse momenta, the positrons emerging from the target need to be immediately focused with a magnetic field tailored for maximum capture efficiency. The special focusing magnet, the so-called matching device, has to provide a strong axial (or azimuthal) magnetic field of the order of few Tesla with the peak close to the target location (ideally at the target surface) and dropping to a constant lower value, typically lower than 0.5~T. When the field is adiabatically decreasing, the system is called Adiabatic Matching Device (AMD)~\cite{helm1962adiabatic,chehab1978second} and, when the transition is abrupt - Quarter Wave Transformer (QWT). The resulting acceptance is a function of the peak and lower values of the magnetic field i.e. $B_0$ and $B_s$, respectively and the capture system aperture $a$ for the particle canonical momenta $p^*_{r0}, p^*_{\phi0}$. This can be obtained, in the case of the AMD as an example, by the condition:

\begin{equation}\label{formula}
\frac{B_0}{B_s} \left(\frac{r_0}{a}\right)^2 + \left( \frac{p^*_{r0}}{\frac{1}{2}e\sqrt{B_0B_s}a} \right)^2 + \left( \frac{p^*_{\phi0}}{\frac{1}{2}eB_sa^2} \right)^2\left[ \frac{B_s}{B_0 \left[\frac{r_0}{a}\right]^2}-1\right] \leq 1,
\end{equation}
where $e = 0.2998$ in the unit [(MeV/$c$)/T/mm] and $r_0$ is the radial distance for a particle's spatial transverse coordinates \cite{Chehab:197428}. It defines the transverse acceptance hyper volume
(acceptance ellipses in the phase space). The positrons with initial conditions ($r_0$, $p^*_{r0}$, $p^*_{\phi0}$) satisfying the Equation~\ref{formula} are considered as the accepted by the AMD system.

\begin{figure}[htbp]
\centering 
\includegraphics[width=\textwidth]{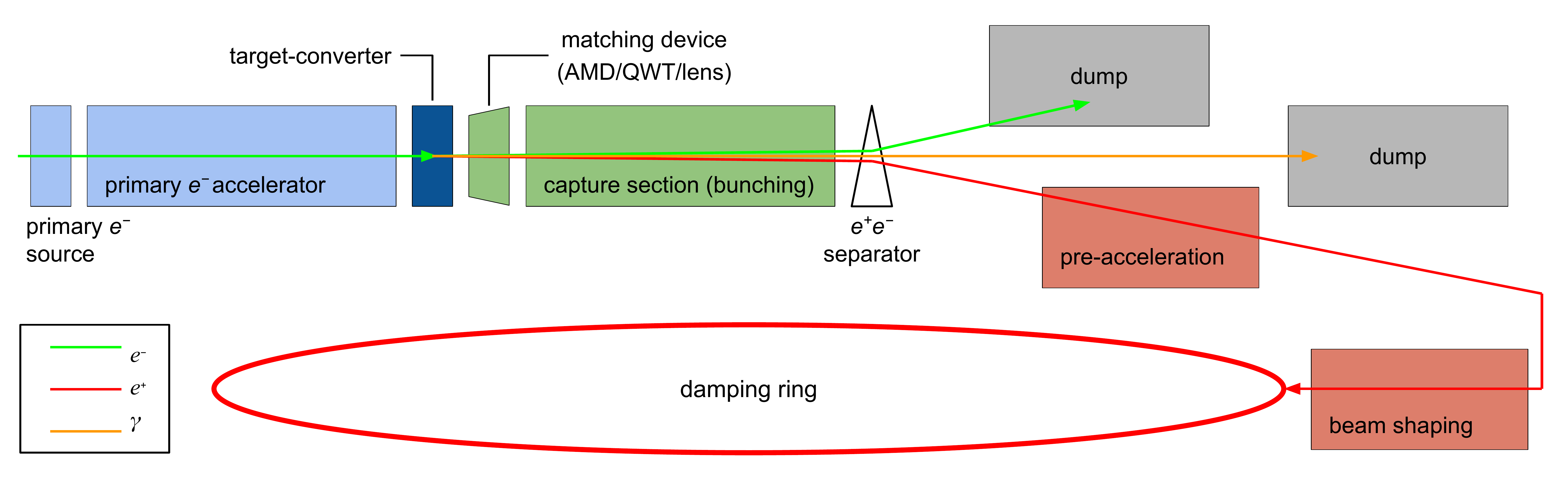}
\caption{\label{fig:main_layout} Different sub-systems of the positron source basic scheme.}
\end{figure}

 Lithium and plasma lenses can be also used as a matching device providing the focusing by azimuthal magnetic field with a focal strength limited by the maximum achievable current in the lens.
 Contrary to the axial fields, the azimuthal magnetic field focuses one kind of particles and defocuses the particles with opposite charge, so discarding the electrons directly in the production stage.
 After the first transverse compression, it is necessary to introduce a longitudinal focusing; the positrons, therefore, enter in a multiple RF accelerating structure line embedded in a DC solenoid magnetic field used in order to bunch and accelerate the beam up to about 200~MeV. The full system composed by the matching device and the RF structures is called the positron capture section~\cite{Chehab:197428}. After this first capture phase, the positrons pass through a quadrupole focusing system (normally a FODO)  and they are accelerated up to the required energy (usually the energy of the Damping Ring). Often Damping Rings (DR) are needed to reduce the positron beam emittances by radiation cooling after the first acceleration stage for achieving their final required values. Before the injection in the DR, energy compressors and diaphragms may provide a 6D~emittance beam shaping to reduce the injection losses. 
 The full positron source system can be illustrated as in Figure~\ref{fig:main_layout}. Eventually, the fraction of the positrons, which is captured for further acceleration and transportation to the Interaction Point, is defined by the capture system and DR acceptances. This introduces the notion, that the positron source efficiency is a global concept starting from the primary beam characteristics and ending with the DR performances.

 This conventional scheme has been used for all circular $e^-e^+$ colliders (ADA \cite{ADA}, ACO \cite{ACO}, DCI\cite{DCI}, SPEAR \cite{SPEAR}, ADONE\cite{ADONE}, VEPP \cite{VEPP}, LEP \cite{LEP_90}, KEKB \cite{Oide:2009zz}, SuperKEKB~\cite{AKAI2018188}, PEP-II\cite{PEP-II, PEP-pos}) and also for the first linear collider SLC \cite{Clendenin:1988np}. 
 In the conventional positron-generation system, a possible scheme to increase the positron intensity is to increase the incident electron beam power (intensity and/or energy). However, the allowable heat load as well as the thermo-mechanical stresses in the target severely limit the power of the primary electron beam. 	
 Thus, the typical positron bunch population obtained from these sources has been a few 1~$\times$~10$^{10}$~ $e^+$/bunch corresponding to average currents less than 1~$\mu$A~\cite{VARIOLA201421}. At present, the positron production rate obtained at the SLC ($\sim $8~$\times$~10$^{10}$~ $e^+$/s) is considered as a world record for the existing accelerators. The main characteristics of the accelerator positron source ever built are listed in the Table~\ref{tab:pos_source_perf}.
 For the future linear colliders, the requested positron production rate is even much higher.
 During the last decade, KEK injector linac has been  upgraded for the SuperKEKB collider project, which requires a 4-fold increase in positron charge compared to previous project KEKB.
 Various studies and new developments, which are of high technological and also engineering importance have been performed to guarantee the required positron source performances. 
 \begin{landscape}
 \begin{table}[p]
 \begin{threeparttable}
 \caption{Performances of ever existed positron sources. Some parameters were not found in literature and, therefore, marked as "--". List of the abbreviations used: Adiabatic Matching Device (AMD), Flux Concentrator (FC), Solenoid(Sol.), Quarter Wave Transformer (QWT), Linac End~(LE), Damping Ring (DR). The target-converter used for VEPP-5 positron source has a truncated-cone shape with the big and small radii being  r= ($\sim$10-->2.5)~mm.}
\label{tab:pos_source_perf} 
\centering
\begin{tabular}{lccccccccc}
\hline 
{\bf Facility} & {\bf SLC } & {\bf SuperKEKB} & {\bf DAFNE } & {\bf BEPCII}& {\bf LIL } & {\bf CESR}& {\bf VEPP-5} & {\bf DCI} \\ 
\hline 
{\bf Research center} & {\bf SLAC} & {\bf KEK} & {\bf LNF} & {\bf IHEP}& {\bf CERN } & {\bf Cornell}& {\bf BINP}& {\bf LAL} \\ 
\hline

Repetition frequency, Hz & 120 & 50 & 50 & 50 & 100 & 60 & 50 &50\\
Primary beam energy, GeV & 30-33 &3.5 &0.19 &0.21 &0.2 &0.15 &0.27 &1  \\
Number of $e^-$ per bunch &$5 \times10^{10}$ &$6.25 \times10^{10}$ &$\sim 1 \times10^{10}$ &$5.4 \times10^{9}$ &$2 \times10^{11}$&$3 \times10^{10}$ &$2 \times10^{10}$ & -- \\
Number of $e^-$ bunches /pulse & 1 & 2 & 1 & 1 & 1 & 7-21 & 1 & 1 \\
Incident $e^-$ beam size, mm & 0.6 & $\sim$0.5  & 1 &1.5 & $\sim$0.5 & 2 & $\sim$0.7 & -- \\ 
Target material &W-26Re &W &W-26Re &W &W &W &Ta &W  \\
Target motion & Moving & Fixed & Fixed & Fixed & Fixed & Fixed & Fixed &Fixed \\
Target thickness/size, mm & 20, r=32 & 14, r=2 & - & 8, r=5 & 7, r= 8 & 7, r=10 & 12, r=($\sim$10-->2.5)& 10.5, r= --  \\ 
Matching device &AMD (FC) &AMD (FC) &AMD (FC) &AMD (FC) &QWT &QWT &AMD (FC) & AMD (Sol.) \\
Matching device field, T &5.5 &3.5 &5 &4.5 &0.83 &0.95 &8.5 (10 max.) & 1.25  \\
Field in solenoid, T &0.5 &0.4 &0.5 &0.5 &0.36 &0.24 &0.5 &0.18 \\
Capture section RF band & S-band& S-band& S-band& S-band& S-band& S-band& S-band& S-band \\
${e^+}$ yield, $N_{e^+}/N_{e^-}$ &0.8-1.2 (@DR) &0.4 (@DR) &0.012(@LE) &0.015(@LE)  &0.006 (@DR) &0.002(@LE) & $\sim$0.014 (@DR)& 0.02 (@LE)  \\
${e^+}$ yield, $N_{e^+}/(N_{e^-} E)$ 1/GeV &0.036 &0.114 &0.063 &0.073 &0.030 &0.013 &0.05 (@DR) &0.02 (@LE)  \\
Positron flux\tnote{*}, $e^+$/s &$\sim 6\times10^{12}$ & $2.5 \times10^{12}$ &$\sim 1 \times10^{10}$ &$4.1 \times10^{9}$ &$1.2 \times10^{11}$ &$7.6 \times10^{10}$ &$1.4 \times10^{10}$&--  \\
Damping Ring energy, GeV &1.19 &1.1 & 0.510 & No  & 0.5 & No & 0.51 & No  \\
DR energy acceptance $\frac{\Delta E}{E}$, \% & $\pm$1 &$\pm$1.5 &$\pm$1.5 & No&$\pm$1 &No & $\pm$1.2 & No  \\

References & \cite{slc-target, clendenin1996compendium} & \cite{prcom1} & \cite{dafne-linac, dafne-linac2}  &\cite{bepc-pos} & \cite{LEP_90, LIL, LIL-robert} & \cite{clendenin1996compendium} & \cite{astrelina2008production}& \cite{prcom2} \\
\hline
\end{tabular}
\begin{tablenotes}\footnotesize
\item[*] Positron flux is calculated from the values listed in the table.
\end{tablenotes}
\end{threeparttable}
\end{table}
\end{landscape} 

\paragraph{Most performing positron source in operation: SuperKEKB.}
\label{sec:SuperKEKB}


SuperKEKB is the $e^-e^+$ collider based on the "nano-beam
scheme" running at the $\Upsilon(4S)$ resonance energy to produce B~meson pairs. The energy of the electron and positron beams are 7~GeV and 4~GeV respectively. This facility is an upgrade of the KEKB collider in order to increase the luminosity to $8\times 10^{35}$~cm$^{-2}$~s$^{-1}$, which is 40 times higher in respect to KEKB~\cite{toge1995kek}. Table~\ref{tab:SuperKEKinj_param} shows the KEKB/SuperKEKB injector parameters. In case of the SuperKEKB, the injector complex includes the 1.1~GeV DR and the related values are given at the final injector energy.

\begin{table}[htbp]
\centering
\begin{threeparttable}
\caption{\label{tab:SuperKEKinj_param} Main parameters of the KEKB/SuperKEKB injectors~\cite{satoh2016commissioning}.}
\smallskip
\begin{tabular}{lrc}
\hline
 &Electrons &Positrons\\
\hline
Beam Energy [GeV] & 8.0 / 7.0 & 3.5 / 4.0\\
Bunch charge [nC] & 1 / 5 & 1 / 4\\
Vertical emittance $\gamma\epsilon_y$ [$\mu$m] & 100/20& 2100/20\\
Horizontal emittance $\gamma\epsilon_x$ [$\mu$m] & 100/50& 2100/100\\
Energy spread [\%] & 0.05/0.08 & 0.125/0.07 \\
Number of bunch & 2/2 &2/2 \\
Repetition rate [Hz] & 50/50 &50/50\tnote{*} \\
\hline
\end{tabular}
\begin{tablenotes}\footnotesize
\item[*] For positron generation a 25~Hz repetition rate operaton mode is used.
\end{tablenotes}
\end{threeparttable}
\end{table}

The requirements on the positron beam imposed by the SuperKEKB collider mainly stem from high storage current (3.6~A) with very short luminosity lifetime (6~min) and small dynamic aperture in the storage ring.
To satisfy the demand from the ring, the new positron source has to provide 4~nC/bunch positron beam, which is four times higher than that of previous positron source for KEKB, with lower emittance of 20~mm$\cdot$mrad. In order to meet the requirements, the positron capture section has been completely reconstructed and a DR has been designed and constructed.

\begin{figure}[htbp]
\centering 
\includegraphics[width=.5\textwidth]{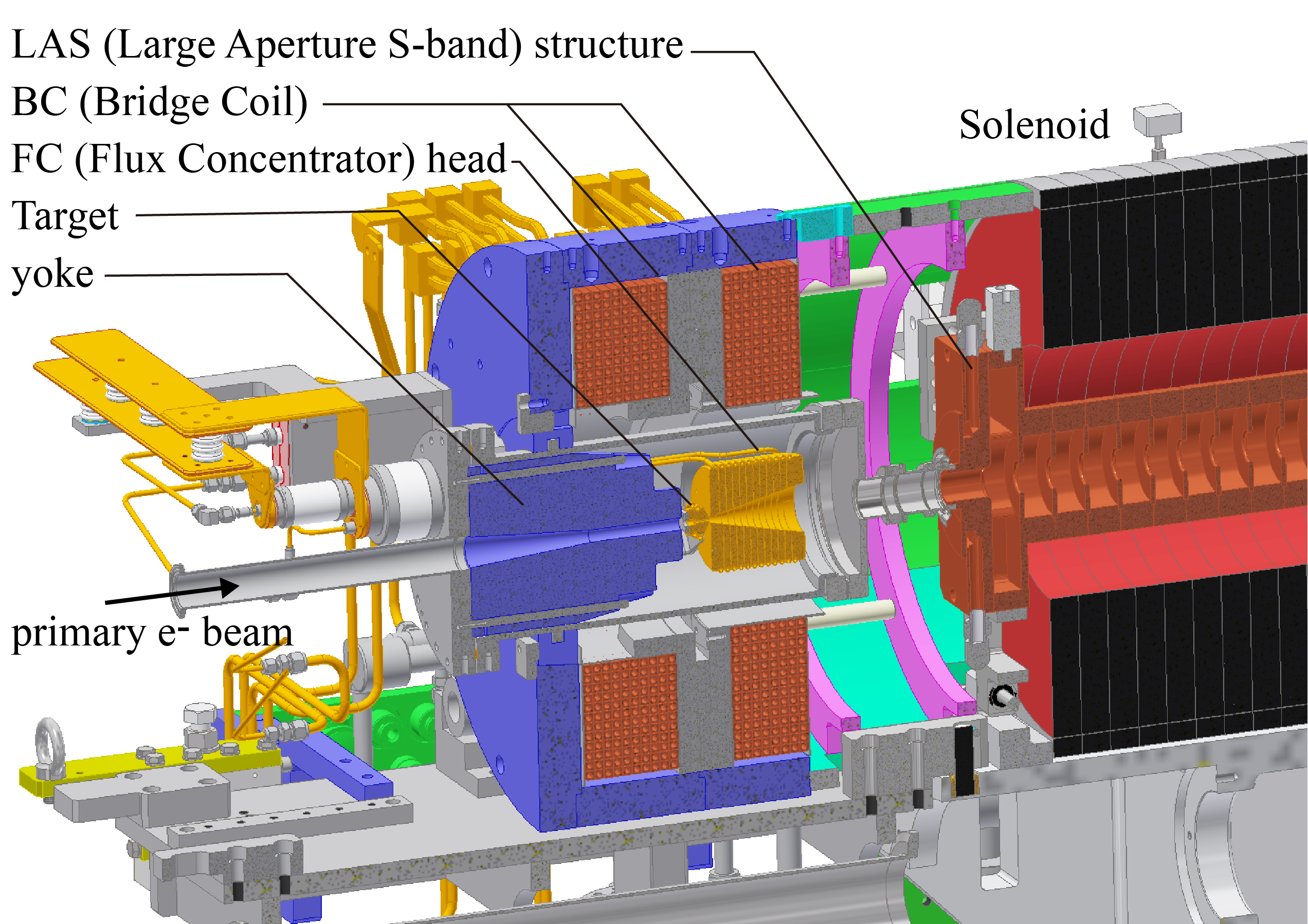}
\includegraphics[width=.4\textwidth]{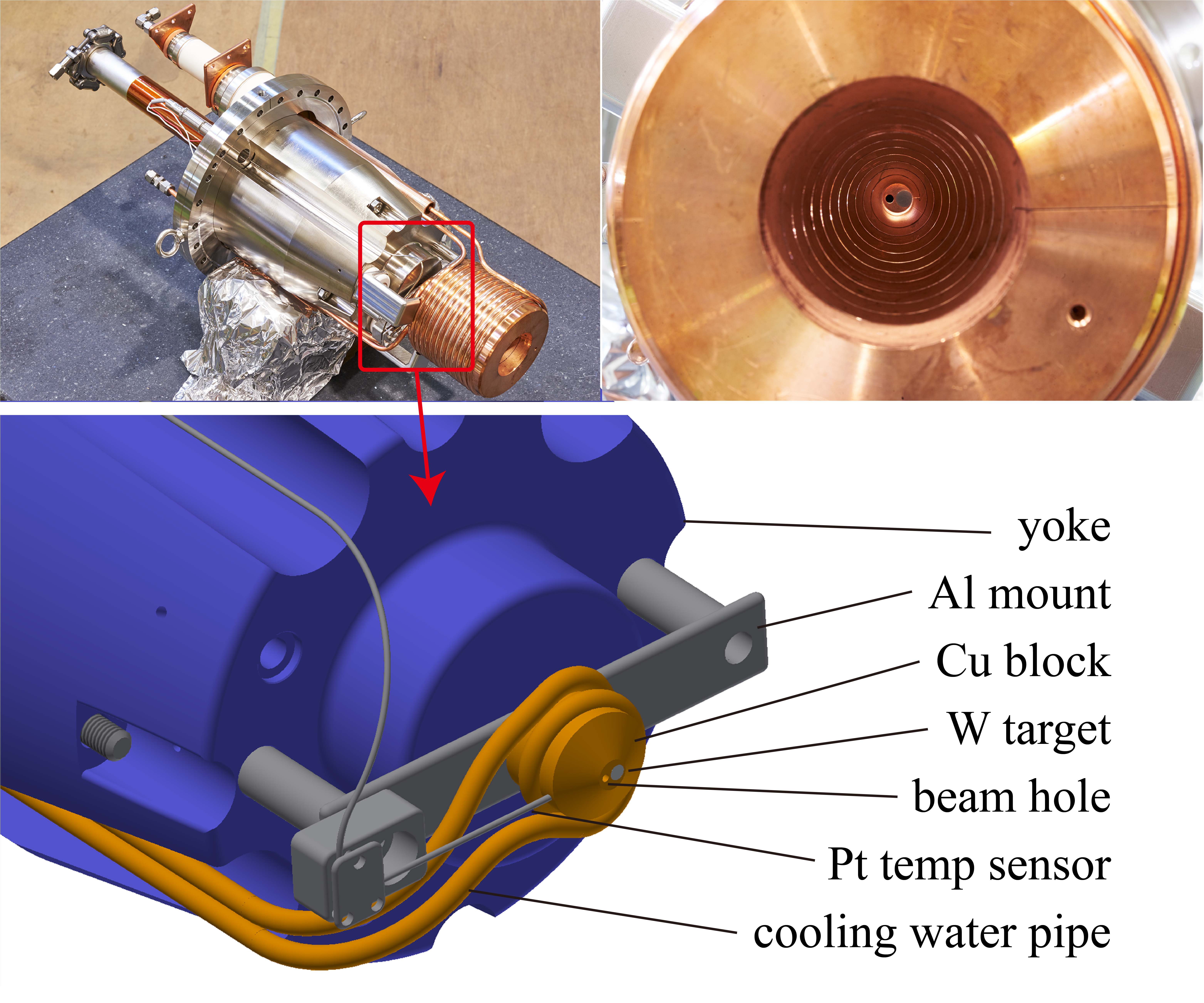}
\caption{\label{fig:superkekeb_cs} Positron source of the SuperKEKB.}
\end{figure}

The SuperKEKB positron sources relies on existing and mature technologies.
Figure~\ref{fig:superkekeb_cs} shows a 14~mm-thick tungsten target-converter and the upgraded positron capture section. It implies a fixed-target design: the low-emittance electron beam passes in the center through a 2~mm hole, while the positron target-converter is installed off-axis. During the operation, the orbits of the electron beam are switched by using two pulse steering magnets upstream of the target.
Systematic studies show that larger hole diameter is preferable for electron beam tuning and emittance preservation. Moreover, as the tungsten target is placed at 3.5~mm from the central axis, it leads to addition degradation of the positron yield~\cite{kamitani2014superkekb}. Therefore, 
recently, the studies have been initiated to explore the possibility to employ the movable/inserting target, which should increase the overall efficiency of the system.

The capture section consists of a Flux Concentrator (FC), a bridge coil, and six large-aperture S-band accelerating structures embedded in the series of the DC solenoid magnets (see Figure~\ref{fig:superkekeb_cs}). 
The positron injector integrates an appropriate 1.1~GeV~DR. Thus,
the positron beam accelerated to 1.1~GeV is injected to the DR for emittance damping.
For this, the energy/bunch compressors are installed in the transfer lines before and after the DR to match the positron beam between the linac and the ring (to avoid additional losses).

Originally, the FC was based on the SLAC-type design~\cite{kulikov1991slc}. The first model designed for the SuperKEKB experienced the severe
discharge problems preventing reaching the nominal field strength of 3.5~T. After the extensive studies and prototyping, a new FC was fabricated with hardening process allowing for high field and stable operation~\cite{enomoto:ipac2021-wepab144}. 
Thus, the FC field strength is 3.5~T at daily operation, with additional 0.5~T generated by the bridge coils. The nominal DC solenoid field strength is 0.4~T.

\begin{figure}[htbp]
\centering 
\includegraphics[width=0.8\textwidth]{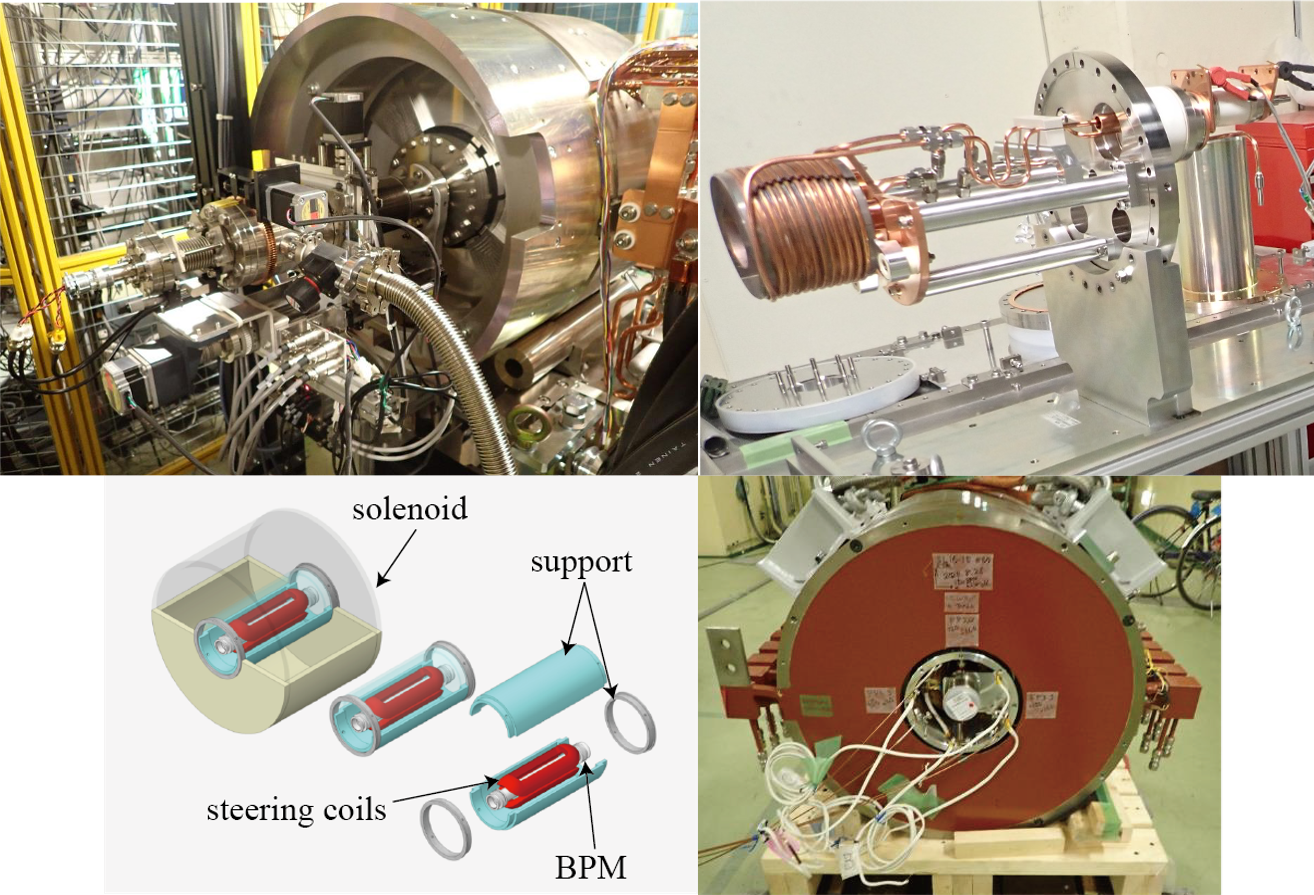}
\caption{\label{fig:SuperKEKBimprov} SuperKEKB positron source R\&D examples: FC testing facility and the fast response BPM/steering coil assembly to be installed in the positron capture section.}
\end{figure}

SuperKEKB covered a large range of positron source/injector investigations and currently rapidly converging to the nominal performance. Table~\ref{tab:SuperKEKBPS_param} summarizes the evolution of the positron beam intensity over the last years.
\begin{table}[htbp]
\begin{threeparttable}
\centering
\caption{\label{tab:SuperKEKBPS_param} The parameters of the SuperKEKB positron source.}
\smallskip
\begin{tabular}{lrcccc}
\hline
& Design & July 2020 & October 2020 & July 2021 & October 2021 \\
\hline
e$^-$ beam energy [GeV] & 3.46 & 3.01 & 2.87 & 2.92 & 2.94\\
e$^-$ bunch charge [nC] & 10 & 8.2 & 8.1 & 9 & 10.3 \\
e$^+$ yield [~$N_{e^+}/N_{e^-}$] & 0.58 & 0.23 & 0.51 & 0.59 & 0.61  \\
e$^+$ bunch charge @CS [nC]\tnote{*} & 5.8 & 1.9 & 4.1 & 5.3 & 6.1 \\
e$^+$ bunch charge @LE [nC] & 4 & 1.3 & 2.1 & 3 & 3.2 \\
\hline
\end{tabular}
\begin{tablenotes}\footnotesize
\item[*] Positron bunch charge at the end of the capture section (CS) or linac end (LE).
\end{tablenotes}
\end{threeparttable}
\end{table}
Such a progress was possible due to numerous achievements, among which the most remarkable are (some of them are illustrated in Figure~\ref{fig:SuperKEKBimprov}):

\begin{itemize}
  \item Improvements of the drive beam parameters. The electron beam intensity of 10~nC can be used for positron generation.
  \item New model of the FC allowing the nominal and stable operation. 
  \item R\&D on the FC led to the availability of the unique versatile test bench with the exchangeable mount for testing the FC. Currently it includes the measurements of the magnetic field, vibrations and temperature distribution.
  \item Development of the very fast response BPM~\cite{suwada2021first}. It allowed monitoring the positron beam in the beginning of the capture section. 
  \item Installation of the steering coils for positron beam inside the capture section.
  \item Improvements on the transport of the positron beam after the DR.
\end{itemize}


Further improvements are mainly focused on increase of the reliability and efficiency of the positron source: mechanical property tests of tungsten material, implementation of the movable target, capture section design (new model of the FC, upgrade of the FC pulsed power supply to reach higher magnetic field, shorten the distance between the FC and the first accelerating structure), improve the transport efficiency between the positron source and the DR, etc..

The positron source of SuperKEKB is the world’s highest intensity positron source currently in operation. 
Despite many challenges, it has demonstrated the reliability of the overall technology and its potential to be used for the various positron source applications.
Much efforts for R\&D and optimization studies for the positron source are currently ongoing at KEK to mitigate limitations to reach the required positron beam intensity. This activity and lessons learned are of great importance for the future lepton collider projects.


\subsection{Future collider projects and associated challenges}
\label{sec:FLC}
Next-generation colliders are required by the HEP and nuclear physics community to investigate the phenomena in the field of Higgs physics and the different scenarios beyond the Standard Model in the post-LHC era. Consequently, the future colliders can be divided in two classes: Higgs factories with center-of-mass energy of $\sim$250~GeV for precision studies of the Higgs boson (circular and linear colliders) and the next energy-frontier machines (linear colliders and Muon Collider project)~\cite{Shiltsev2021}.
Hence, given the past experience, it is crucial to explore the TeV scale with the colliders based on the $e^+e^-$, $h/e^+/e^-$ and, certainly, $hh$ collisions.
As far as the lepton-based collisions are concerned, the most feasible options currently under consideration are $e^+e^-$ linear colliders (ILC and CLIC) and $e^+e^-$ circular colliders i.e. FCC-ee and CEPC, whereas the $\mu^+\mu^-$ Muon Collider is not yet a mature option being in the conceptual design phase. The full physics program of a future Large Hadron electron Collider (LHeC) requires both $pe^+$ and $pe^-$ collisions~\cite{LHCE2012}.
 Each of these collider options has its own advantages, drawbacks and challenges, however, a requirement for very high luminosity leads to the fundamental role played by the particle injectors and positron source in particular.

In addition, the necessity of using both polarized electron and positron beam at the future colliders is well established and has been comprehensively analyzed in~\cite{moortgat2008polarized}. 
It presents additional degree of complexity in original design of positron sources, requiring production of polarized photons followed by polarization transfer. There are three methods to generate circularly polarized high-energy (>10’s MeV) photons, which can be converted to the longitudinally polarized positrons. Radiation from helical undulator and Compton scattering-based sources are intended for high energy colliders, and bremsstrahlung from polarized electrons is also considered at lower energies typical for Hadron Physics (MAMI, MESA, JLab), and also for the much smaller energy range of Atomic Physics and Material Science~\cite{grames2018polarized}. In these domains, a conventional positron source design using initially polarized electrons and capturing high-energy positrons has been demonstrated to be particularly efficient~\cite{PhysRevLett.116.214801}.

Table~\ref{tab:pos_source_future} shows the challenges of some future colliders, especially the positron source related parameters. In such a way, the intensities required from the positron sources at the future colliders CLIC, ILC or LHeC are a few orders of magnitude higher (up to $\sim10^{15}$ $e^+$/s at the LHeC) than that delivered by ever existed facilities \cite{Nature}.
For the FCC-ee, a positron bunch intensity of 2.1~$\times$~10$^{10}$ particles (3.35~nC) is required at the injection into the pre-booster ring allowing for a positron accepted yield of 0.5~$N_{e^+}/N_{e^-}$ if safety margins are neglected and the maximum allowed electron drive beam current is used~\cite{FCC_CDR_2}. This value is comparable with the positron yield envisaged at the SuperKEKB. The positron flux, however, is estimated to be similar to the flux obtained at the SLC (see Table~\ref{tab:pos_source_perf}). 
 
 Intense positron sources may play also an important role for even higher-luminosity colliders and, in particular, for positron-based muon production (cf. LEMMA proposal)~\cite{alesini2019positron}. LEMMA is an alternative scheme to produce the muon beams using positrons impinging on a target at the muon production threshold. For this reason, deep studies of a positron source capable to deliver the exceptional flux of about $\sim10^{16}$ $e^+$/s are needed to narrow down the possible design choices and define the R\&D directions to mitigate the critical issues. Creating low-emittance energetic muon beams would open the door to a new generation of lepton colliders, which was also stated by the recent EU HEP Strategy update~\cite{adolphsen2022european}. 
\begin{table}[htbp]
\begin{threeparttable}
     \caption{Future Positron Collider Projects~\cite{Zhao:2735292, riemann2020updated, resultsILCconv, zimmermann:ipac12-weppr076, Alesini:2019tlf,Wang:2017mjk, FCC_CDR_2, Shiltsev2021}. \label{tab:pos_source_future}}
     \begin{tabular}{lcccccc} \hline
         Project                       & CLIC & ILC   & LHeC (pulsed) & LEMMA & CEPC &  FCC-ee  \\ \hline
         Final e$^+$ energy [GeV]      & 190 & 125    &  140       &    45        & 45       & 45.6         \\ 
         Primary e$^-$ energy [GeV]    & 5    & 128** (3*)  &  10       &      --      &   4    & 6         \\ 
         Number of bunches per pulse             & 352  & 1312 (66*) &    $10{^5}$           &   1000         & 1      & 2        \\ 
     Required charge [10$^{10}$ e$^+$/bunch]  & 0.4  & 3 & 0.18 & 50 & 0.6 & 2.1 \\
     Horizontal emittance $\gamma\epsilon_x$ [$\mu$m]  & 0.9  & 5 & 100 & -- & 16 & 24 \\
     Vertical emittance $\gamma\epsilon_y$ [$\mu$m]  & 0.03  & 0.035 & 100 & -- & 0.14 & 0.09 \\
         Repetition rate [Hz]          & 50   & 5 (300*)   &      10         &     20       & 50      & 200         \\ 
         e$^+$ flux   [10$^{14}$ e$^+$/second] & 1    & 2    &      18         &   10--100         &   0.003    & 0.06         \\ 
         Polarization & No/Yes*** & Yes/(No*) & Yes & No & No & No    \\ \hline

  \end{tabular}
\begin{tablenotes}\footnotesize
\item[*] The parameters are given for the electron-driven positron source being under consideration.
\item[**] Electron beam energy at the end of the main electron linac taking into account the looses in the undulator. 
\item[***]  Polarization is considered as an upgrade option.
\end{tablenotes}
\end{threeparttable}
\end{table}

The choice of collider type/technology, thus, greatly influences the design of the positron source and its performance. The main challenges remain the same: the requested high intensity/current, low emittance, polarization and high reliability.
At linear colliders, the positron source should be designed to produce all the current necessary for the collider at one shot.  So the different designs foresee multi-bunch operation either in CW or pulsed mode resulting in an average current at least more than one order of magnitude higher than the current state of the art (see Table~\ref{tab:pos_source_future}). On the other side, the situation is less demanding for the circular colliders, where the positron bunches can be stacked in the collider followed by the  top-up operation.
As far as polarization is concerned, it is highly requested by the linear collider projects, while still under discussion for the circular options.



\paragraph{Novel approach: photon-driven positron sources and polarized
bremsstrahlung.}

A common feature of the future collider projects  is the requirement  to generate the high peak and/or average current beams and, as far as possible, with an important degree of polarization for both. 
In this context, the intensity required for the positron source being up to $\sim10^{16}$~$e^+$/second (see Table~\ref{tab:pos_source_future}) is not within the reach of proven technology and concepts.
It would be difficult to realize by the conventional scheme using electrons, which are converted into the positrons because of the high heat load inside the target as well as a thermal stresses/shocks caused by the inhomogeneity of the primary beam energy deposition. 
Thus, the target thermo-mechanical stresses given by the PEDD and the average heating impose a physical limit on the peak and average positron current.
On the other hand, this method is also not attractive due to the lack of positron polarization unless a polarized electron beam is used to generate the positrons.\footnote{In this case, 
the polarization transfer from the electron to the bremsstrahlung photon takes place that results into the polarized positron production. This method of positron production usually is called polarized bremsstrahlung ~\cite{PhysRevLett.116.214801}.}

\begin{figure}[htbp]
\centering
\includegraphics[width=1.\linewidth]{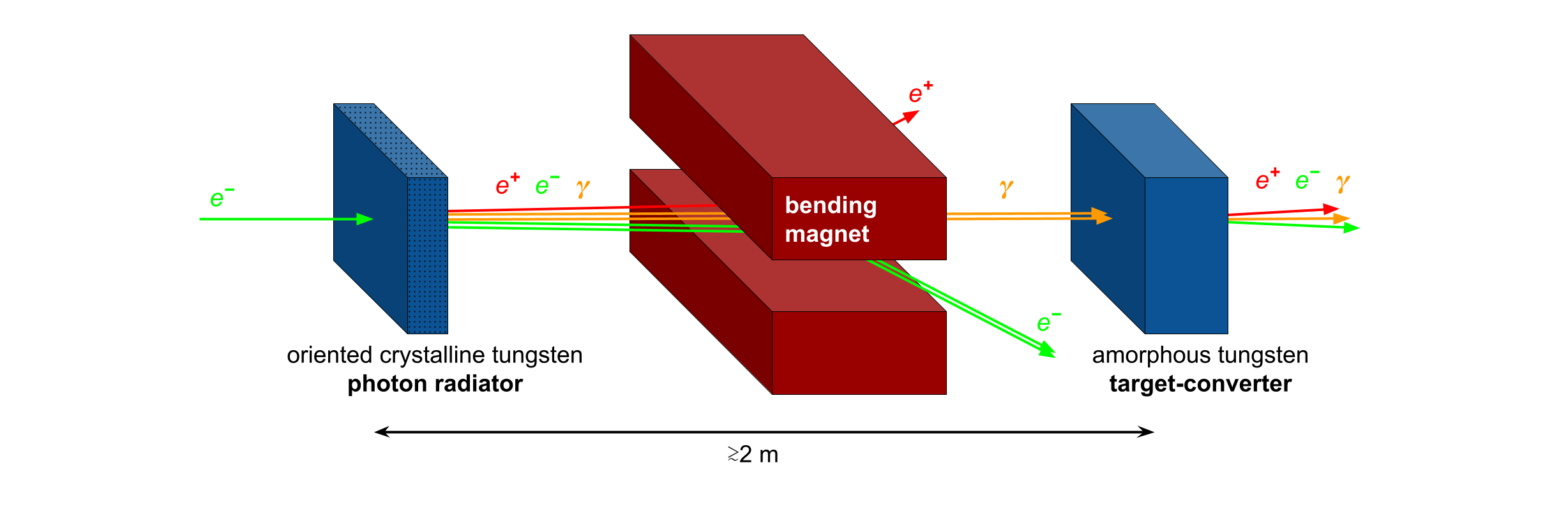}
\caption{Hybrid scheme for the unpolarized positron production proposed to be used for the future colliders. Simple two-stage version, which includes a crystalline photon radiator followed by an amorphous converter. A bending magnet is placed between the targets to redirect the charged particles emerging from the crystal target.}
\label{fig:sources_twoStage}
\end{figure}

To overcome the very challenging constraints imposed by the target (mainly thermo-mechanical), a better solution of using a two-stage process was proposed for the positron production. Initially it was proposed to be used at the future liner colliders and later at other projects. The first stage is a generation of photons/gamma rays. In the second stage the electron and photon beams are separated and the latter is sent to the target, where the photons are converted into the $e^-e^+$ pairs. This strongly reduces the large ionization energy losses given by the low energy population of the drive beam leading to the heating of the target.  A major difference now between the different schemes to produce positrons is how these photons are generated. An example of one such positron source using channeling radiation in oriented crystals acting as an intense source of the photons is illustrated in Figure~\ref{fig:sources_twoStage}.

\paragraph{Undulator-based polarized positron sources.}

The usage of undulator radiation allows obtaining both polarized positrons and electrons has been proposed~\cite{balakinconversion, Balakin:1988se} and studied in~\cite{mikhailichenko1925conversion}. The radiation from helical undulator is circularly polarized. This makes it especially attractive in the context of polarized positron sources. Since there is a well-defined correlation between the photon energy, its emission angle and polarization, by performing an energy selection of the undulator photons, one can choose a desirable degree of photon polarization and so, the positron beam polarization.

For a typical size of the undulator period being $\lambda_u$~=~1~cm, one needs an electron beam energy of $\sim$150~GeV to get $\sim$10~MeV photons useful to create $e^-e^+$ pairs in the target-converter. This leads to a need of very high electron beam energy to drive such a facility; this is available in HEP collider projects but it introduces a correlation between electron and positron lines imposing constraints to be taken into account in the design phase. For fixed magnetic field and undulator period, the number of photons generated is proportional to the undulator length.
In this context, this innovative undulator-based positron source driven by the main electron beam is conceived as part of ILC baseline design~\cite{ILC2013TDRvol3Accelerator}.

The ILC TDR concentrates on a baseline machine of 500~GeV centre-of-mass energy. However, the discovery of a Higgs boson with a mass of 125 GeV opens up the possibility of reducing cost by starting at a centre-of-mass energy of 250~GeV. The ILC is now proposed with a staged machine design, with the first stage at 250~GeV with a luminosity goal of $2 ab^{-1}$ (total length of approximately 20~km). Recently, the physics motivation, the accelerator design, the run plan, the proposed detectors and physics to be studied have been reviewed~\cite{Bambade2019ilc}.

\begin{figure}[htbp]
\centering
\includegraphics[width=1.\linewidth]{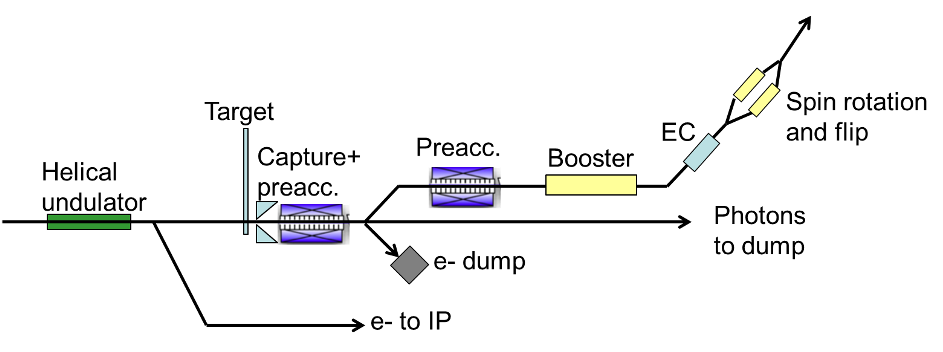}
\caption{A schematic layout of the ILC undulator-based polarized positron source. Courtesy of ILC positron source collaboration.}
\label{fig:undulator}
\end{figure}

The undulator-based positron source is located at the end of the main electron linac. It consists of the superconducting helical undulator to generate the circularly polarized photons, the thin target-converter made of Ti6Al4V alloy, the capture section, positron pre-accelerator, energy compressor, spin flipper and spin rotation (see Figure~\ref{fig:undulator}). The nominal design should ensure a positron yield of 1.5~$N_{e^+}/N_{e^-}$ at the 6~GeV DR. Among the main challenges of this proposal are a high number of positrons to be generated ($\sim3 \times 10^{10}$ e$^+$/bunch, 1312~bunches/pulse at 5~Hz repetition rate) and wide energy range of the primary electron beam (from 128~GeV in the current baseline to 500~GeV in case of the 1~TeV ILC upgrade). 

A proof of principle experiment E--166~\cite{alexander2008observation} has been performed in the FFTB at SLAC to demonstrate production of polarized positrons for a further implementation in the undulator scheme at the ILC~\cite{Alexander:2009nb}. 
Some time ago, a fully working high-field, short-period $4$-m-long prototype of the SC helical undulator cryomodule suitable for use in the future ILC positron source has been designed, manufactured and successfully tested~\cite{Scott2011Undulator}. The required on-axis peak field of 0.86~T has been achieved with $\lambda_u~=~11.5$~mm and beam aperture of 5.85~mm.  

In order to achieve the required intensity of positrons the 231-m-long undulator with a period of 11.5~mm and a variable undulator strength parameter K~$\le$~0.92 is considered. The final positron beam polarization is estimated to be~$\sim$30\%. The generated photon beam with an average power of 72~kW has an opening angle of approximately 3~$\mu$rad owing to high energy of electron beam. 
This leads to a very challenging design of the photon dump and the collimator system~\cite{Alharbi2021ipac}. Technical solutions for the photon dump, based either on water or graphite are being considered.
Due to the high value of the PEDD, the target-converter is conceived as a 1-m wheel spinning in vacuum with 2000~rpm (100~m/s).
On the other hand, the relatively small total deposited power (2.2~kW for the 250~GeV collider operation mode) in the 7~mm thick Ti6Al4V target allows designing the target cooled by radiation only~\cite{riemann2020updated}. 
In the ILC TDR the QWT was chosen as a matching device installed downstream the target at electron beam energies above 150~GeV. The QWT with peak field of 1.04~T has more stable magnetic field and bigger aperture (r = 11~mm) compared to considered earlier pulsed FC (3.2~T and r = 6~mm). However, the obtained positron yield with QWT-based capture system at 128~GeV electron beam energy is too low being 0.9~$N_{e^+}/N_{e^-}$. Recently a 3-5~T pulsed tapered solenoid was proposed to increase the positron yield to 1.5~$N_{e^+}/N_{e^-}$~\cite{riemann2020updated}.

To mitigate the risks associated with the ILC project, an electron-driven conventional positron source is being developed as a backup system for the positron source. The progress of the electron-driven positron source for the ILC was recently reported including the feasibility studies in terms of the heat\&stress of the accelerator components, the positron yield estimate for realistic conditions such as beam loading effect as well as the status of the target system R\&D~\cite{Nagoshi2020nima}.
In such a way, nowadays, the ILC positron source (both undulator- and the electron-driven) remains the most mature positron source proposed for the future collider in terms of pre-design studies and prototyping.

\paragraph{Compton scattering-based polarized positron sources.}

Another attractive and compact solution foresees the use of an high-power laser beam and the Inverse Compton Scattering (ICS) interaction to create such photons~\cite{bessonov1996method, okugi1996proposed}.
The electron beam requirements in this case are greatly reduced while still reaching higher photon energies. Considering the scattering with a typical laser ($\lambda$~=~515~nm), the electron energy required to generate 30~MeV photons is around 1.0~GeV and very small spot sizes have to be maintained only over relatively short interaction lengths (less than few cms). 
In 2005, a proof of principle experiment for the Compton scattering-based scheme for polarized positron generation was performed at the KEK Accelerator Test Facility (ATF)~\cite{PhysRevLett.96.114801}.

Several options for the future linear collider positron source based on
Compton scattering have been proposed~\cite{rinolfi2009clic}. Today, they can be classified according to the electron source used for the Compton scattering: the linac scheme, the storage ring scheme or so-called Compton Ring and the energy recovery linac scheme.
For all of them, the polarized positron current produced is not sufficient to fulfill the future linear collider requirements. Therefore, the application of the multiple-point collision line and multiple stacking of the positron bunches in the DR were investigated~\cite{chaikovska2012polarized, zimmermann2009stacking}. 

On the other hand, owing to the small size of the Compton (Thompson) cross section, the demands of such solution on the high-power laser system are extremely challenging\footnote{In case of Gamma Factory proposed at CERN~\cite{2019gamma}, which uses partially stripped ion beams and their resonant interactions with laser light, the resonant photon absorption cross section can be up to a factor $10^9$ higher than for the ICS of photon on point-like electrons. The proof of principle experiment was already proposed~\cite{Krasny:2690736}.}. 
The time format of ILC beams, for example, is constituted by an elevate number of bunches ($>$1000) per RF macropulse, with macropulse repetition rates of 5-10~Hz. At visible wavelengths, Joules of energy are required in order to provide sufficient photon density for the generation of one photon per incoming electron in the laser-beam interactions. The laser system should, therefore, provide multi-MW-class average power within a burst mode matching the electron bunch time-format. Using the additional degree of freedom offered by fast kickers, one can imagine to reformat the positron source to 30~KHz repetition rate and recreate the collider bunch format only after the DR, easing somewhat the peak and average requirements on the laser. Notwithstanding the exceptional progress of the RF and of the laser technology in the last decades, even this latter kind of laser system does not exist yet.
Various new concepts, such as stacking cavities and optical energy re-circulation have been proposed to address the lack of a suitable laser source for this application~\cite{chaikovska2016high, PhysRevAccelBeams22093501, PhysRevAccelBeams.23.051301, Amoudry:20, PhysRevAccelBeams21121601B}.

In Murokh et al.~\cite{rbtoptical} the authors present
an alternative approach for an independent high-current polarized positron source based on the combination of laser-based acceleration with the observation, that the electron and laser beams are only minimally degraded in a ICS interaction. The laser pulse can then be used not only to drive the Compton scattering process, but also to accelerate the electrons to the required GeV-level for energetic polarized photon production. At the same time, after the ICS interaction point, the kinetic energy stored in the electrons can be recuperated with an high efficiency Free-Electron Laser (FEL) amplifier operating in the Tapering Enhanced Stimulated Superradiant Amplification (TESSA)  regime to replenish the laser pulse before it is redirected to scatter against the next electron bunch. Due to the limited electron beam and laser power requirements of this scheme, the electron current used in the accelerator can be very large and, even with the yield of 0.1~$N_{e^+}/N_{e^-}$ after conversion of the gamma-rays in the target, positron fluxes of up to 10$^{15}$ $e^+$/s could be achieved. 

It should be emphasized that due to a common technological constraint of all the above mentioned schemes being the average laser power of the optical systems, the Compton scattering-based polarized positron sources are considered only as the alternative solutions for the future collider projects.
Presently, it is proposed as a preferred option for an upgrade of the CLIC positron source~\cite{CLICcdr}.

\paragraph{Crystal-assisted positron sources.}

The basic and important parameter for a positron source is the quantity of photons converted in the $e^-e^+$ pairs. The larger this number,  
the larger the number of positrons produced. In conventional scheme, bremsstrahlung radiation is the main source of photons. In order to get a large number of photons the electromagnetic shower in the target needs enough thickness (or radiation lengths) to expand. It is known~\cite{katkov1998electromagnetic}, however, that axially aligned crystals are much better radiators than the conventional amorphous converters. Electrons penetrating the crystal at glancing angles to the axes or planes are channeled and emit channeling radiation. This radiation is providing a large number of softer photons with respect to bremsstrahlung~\cite{chehab1989study}. The effective radiation length for a tungsten W~crystal aligned on its <111> axis and hit by a 8~GeV electron beam is only 0.6~mm, whereas the radiation length associated to the bremsstrahlung for the amorphous target is 3.5~mm. The choice of crystalline radiators leads, consequently, to a reduced thickness of the conversion length for the same number of positrons. Radiation and conversion may occur in the same crystal target and provide high positron yields as observed in the WA103 experiment at CERN~\cite{CHEHAB200241, ARTRU2005762}. However to avoid important energy deposition in the crystals, which may deteriorate the crystal qualities, it seemed better to separate two functions: radiation and conversion. 

In this context, investigations performed at Laboratoire de L'accélérateur Linéaire (LAL) and Institut de Physique Nucléaire de Lyon (IPNL) in connection with a theoretical group of V.~N.~Baier of BINP led to a proposal of a hybrid scheme based upon a relatively new positron source scheme. It uses, as a primary beam, an intense photon pulse produced by high-energy (some GeV) electron bunch channelled along a crystal axis (channelling radiation). Several experiments at CERN, KEK and Orsay, have been performed to investigate such a possibility~\cite{artru1996axial, CHEHAB200241,PhysRevE.67.016502}. They have shown very promising results for the enhancement of the positron yield and the reduction of the energy deposition in the target, if compared to the conventional one, especially, applying the hybrid scheme, where the photon and the positron production are physically separated~\cite{ARTRU20083868}.
In this approach, a bending magnet can be installed between the radiator and the target-converter, which allows sweeping off the charged particles emerging from the crystal (Chehab - Strakhovenko - Variola scheme), hence reducing the thermal and radiation load being of great importance for the linear collider projects.
Moreover, a new option of the target-converter can be also considered implying the use of a granular target made of small spheres providing better heat dissipation and better resistance to thermal shocks~\cite{cheng2012positron}. The hybrid scheme, thus, has been adopted by CLIC as a baseline for the unpolarized positron source. 
The hybrid scheme is currently being also investigated in application to the circular colliders (FCC-ee) requiring relatively moderate beam intensities~\cite{Chaikovska:2019ztn}.  

\paragraph{The polarization for low-intensity CW beams: polarized
bremsstrahlung.}
The polarized bremsstrahlung~\cite{PhysRevLett.116.214801} as a method to produce polarized positrons is considered to be not very efficient in a framework of the future linear colliders. It is mainly due to the low positron capture efficiency at high energy part of the positron spectrum, where the population characterized by a high degree of polarization is located. However, in 2018, the Positron Working Group at Jefferson Lab, with over 250~members from 75~institutions submitted a Letter of Intent titled Physics with Positron Beams at Jefferson Lab 12~GeV, promoting a series of experiments using positron beams that could occupy CEBAF operations for more than 3 years~\cite{Afanasev2019physics}. Recently, two proposals focusing on Deeply Virtual Compton Scattering (DVCS) were conditionally approved based upon the availability of >100 nA polarized or >1~$\mu$A unpolarized beam currents, respectively.
A conceptual design study is underway to develop CW positron beams for CEBAF. CW electron beam with polarization >85\% will be used for  generation of polarized positrons.
Currently, three different electron beam energies are considered: i) 10~MeV, ii) 123 ~eV and iii) 1090~MeV~\cite{Arrington2021physics}.
10~MeV option is below the photo-neutron production threshold of most materials, however, suffers from the lowest yield and highest demand on the polarized electron source.
The option iii) would provide a significantly higher positron yield, but likely has the largest radiological and construction footprint.
The option ii) is a better compromise between positron yield and footprint.
The 1~mA electron beam with an energy of 123~MeV passing through a 4-mm-thick single tungsten target-converter\footnote{This is the optimum thickness for the maximal product of positron yield times positron polarization squared.} will deposit approximately 17~kW in the target. The design studies of rotated solid targets and liquid-metal jet targets are presently being considered.
The engineering study for a 2-3~Tesla DC solenoid matching device, the  positron capture section and energy-selection chicane are currently underway.
Hence, the final positron source design should be capable to provide at least 10~nA current and >60\% of the positron beam polarization. 



\section{Physics and technology challenges to improve the future positron source performances}
Positron sources are complex devices, where each subsystem (production, capture, acceleration, and DR injection) has an impact on the final performance of the positron source.
The required beam intensities and emittances (see Table~\ref{tab:pos_source_future}) impose technological challenges for the positron source design (target technology, cooling systems, capture optics, power dissipated on the structures, remote handling/target removal engineering design, etc.). The complete optimization of the positron production requires not only maximizing the total polarized or unpolarized positron yield, but also innovative studies of target thermodynamics which influences the performance of the positron source. 
In this context, investigations/studies of the heat dissipation and thermo-mechanical stresses in the targets and closest beamline components, technological R\&D and experimental testing of the innovative targets and positron capture systems are all mandatory for more robust and reliable positron source designs to meet future needs. 

Any increase of positron production and capture efficiency reduces the cost and complexity of the driver linac, the heat and radiation load of the target-converter system and increases the operational margin. The latter proved to be a very critical issue for the operation of the positron sources of the SLC and KEKB/SuperKEKB, the most performant accelerator-based positron sources built so far.

\subsection{Targets: thermo-mechanical and radiation problems}
Different effects occur in materials under irradiation depending on the parameters of the incoming beam (its energy, charge, dimensions and time structure) and, certainly, the material properties~\cite{Bertarelli:2016obw, Mokhov:2016fdl}. 
In general, the positrons are produced by impinging electron beams onto a high-Z material.
The energy deposition density, induced by the incident electrons inside the target, is strongly non-uniform in r-(radial) and z-(longitudinal) direction. This causes a rapid, adiabatic rise of temperature, according to the duration of the electron pulses, in general in the $\mu$s range. This temperature profile persists over some ms before it becomes more uniform by thermal diffusion. This has been theoretically modeled in~\cite{cesarini2021theoretical}.
During this time, quasi-stationary thermal stresses will occur, which are compressive along the axis of the target and tensile at the outer boundary of the target. And this, due to thermal cyclic loads can lead to fatigue and, finally, to failure of the target material~\cite{Bharadwaj2001slc}. Large beam sizes and, thus, large targets would diminish such effects. 
In the past, the breakdown of the SLC target led to the systematic but empiric determination of the acceptable PEDD. The studies undertaken at LANL and LLNL~\cite{maloy2001slc, stein2001thermal} concerning the SLC target showed that the PEDD of 35~J/g might not be exceeded. This is now considered as a maximum tolerated value of the PEDD in the tungsten targets for a positron source design.

As stated above, photon-driven positron sources, like the hybrid scheme or the undulator-driven scheme (proposed for the ILC) have the advantage of depositing less beam energy in thinner target and still providing sufficient positron yields. This reduces also the technical complexity and the handling of the target station. 
For below $\mu$s beam pulses, thermal shock effects and/or vibrations in the target material may occur. These are driven by inertia of the target material, since it cannot thermally expand as fast as it is heated. As a guide line, such effects appear, when the transition time $\sigma$/$c$ ($\sigma$ being the radial PEDD profile extension inside the cylindrical target rod and $c$ is the velocity of sound of the target material) becomes longer than the pulse duration of the impinging beam pulse~\cite{sievers2003stationary}. 
As an example, a typical 1~mm of radial PEDD profile and a velocity of sound of 4~km/s would lead to a transition time of 250~ns. These effects have been observed in many laboratories (SLAC, CERN, KEK, etc.), and have led to warping, bending of slender target rods or to radial cracks induced by vibrations. These beam induced heating and shocks may also affect components, like the AMD, placed very close to the target and must be taken into account. In general, to minimize such effects, for a given average beam power, an high pulse frequency with long pulse duration would be desirable. 
Nevertheless, it appears that a real modeling and computational effort on the transient regime, taking into account pressure waves propagation and mechanical impedance matching in the support is actually missing. This should be an important improvement to have a deep understanding of the PEDD failure mechanism and to find eventual cures.

To maintain a low average temperature of the target, efficient cooling to evacuate the average power from the target, is required. For slender targets, cooling by radiation and aided by natural or forced Helium gas cooling, has been employed. However, for efficient cooling by radiation, high temperatures of above 1000$^{\circ}C$  may have to be tolerated.
For laterally large targets, the heat can be evacuated radially by heat conduction into a water circuit (used or foreseen for the positron sources at SLAC, KEK/SuperKEKB, ILC, etc.).
For high-intensity positron sources, the average power as well as the cyclic thermal loads may be too high to be tolerated by a stationary target. Therefore, moving, rotating (ILC positron source) or tumbling/trolling wheel (SLC positron source), water or radiation cooled targets have been or are envisaged to be employed. This helps to spread the heat as well as the radiation damage over a larger area.

Engineering solutions have been developed or are under consideration, which, however, have to be compatible with the adjacent positron capture system. 
For very high positron yields, which are contemplated  for the far future, like muon colliders, liquid metals (liquid Lead or Eutectics, being liquid at room temperature) as targets may be used. Technology, in particular from reactor technology or from  spallation sources,  exists  or can be adapted. Using liquid metals, driven through the incident beam in a closed loop will spread the heat input and the radiation damage over the total volume of the liquid metal circulating in the loop, while the cooling occurs in a heat exchanger, external to the beam area. 
The principal constraint is given by the possible cavitation of the liquid, due to the beam high-energy deposition, with the consequent loss of the static liquid flow regime. 
The issue of damaging the container by cavitation, induced by short and intense particle pulses, has been experienced and ways to mitigate it have been found. To bypass the problem of the containment of liquid metal targets, metal jets and free falling curtains or droplets have been proposed for neutrino factories.
Suitable simulation codes are available to study fluid mechanics and any dynamic effects or shocks in such systems.

Another target damage mechanism, related to irradiation, is radiation damage~\cite{Artru:359257}. Radiation damage of the target and other beamline component materials manifests itself in atomic displacements causing the structural material damage. The widely used measure of the material deterioration is the number of displacements per atom (dpa), which depends on the impinging particle type, its characteristics as well as a material properties. Systematic evaluation of the dpa is now essential element of the positron source design being important for the long-term reliable operation.
  
Overall, the high-power targets and their reliability under the severe operation conditions are of interest in several fields (neutrino physics, muon physics, neutron physics, etc.). One should profit from the mutual experience, studies and, eventually, try to join the effort for the future developments.
It is important to stress, that the experimental tests are of great importance for both very high values of the instantaneous power load (stresses and shocks) and target lifetime studies.
Assessment of the failure criteria in materials under irradiation via thermo-mechanical stresses, thermal shocks (occurring in the $\mu$s scale) and  longer-term phenomena (radiation damage, fatigue) to predict the target and any component damage remains one of the major challenge for the current and future positron source designs.


\subsection{Capture section technology}
The positrons and electrons produced in the target-converter are captured and accelerated up to a few hundreds of MeV in the capture section before being separated by a chicane (see Figure~\ref{fig:main_layout}) and accelerated in the pre-accelerator to be injected in the DR.
Thus, another way to improve the positron source performance is to increase the capture efficiency. This could also mitigate the target thermal load, which is the critical aspect of the positron source design, especially at high intensities. 

The capture section is composed of the matching device and capture linac. The main role of the matching device in the capture system is to reduce the large angular divergence of positrons after the target by increasing the beam radius. In this way, the positron beam occupies all the geometrical acceptance at the entrance of the accelerating structures. After this, the whole capture linac is encapsulated inside a DC solenoid. The typical magnetic field values of such solenoids used up to now are around 0.2--0.5~T, however, higher values are required to improve the capture efficiency. It is employed to focus the positrons and avoid the positron losses, caused by the high divergence of the positron beam at the beginning of the capture section, while the RF accelerating field provides the longitudinal compression.

At present, the most used matching device technologies are the AMD realized by means of the FC and the QWT made of the DC or pulsed coils. 
The advantage of using the AMD rather than the QWT as a matching device, is that it allows increasing the accepted positron yield by capturing positrons within a wide energy band. Nevertheless, the drawback of the AMD is a lengthening of the bunch, which is afterwards responsible for the increase of the energy spread in the capture linac.
These systems have been successfully used at SLAC~\cite{bulos1985design},  LAL in Orsay~\cite{chehab1983adiabatic} and being currently in operation in SuperKEKB at KEK (see Table~\ref{tab:pos_source_perf}).
For capture linac, the normal-conducting S-band accelerating structures were in use for all the existing positron sources (see Table~\ref{tab:pos_source_perf}). In general, large-gradient and large-aperture accelerating structures are required to ensure sufficient longitudinal
and transverse acceptance for the positron beams. Therefore, the L-band option is currently under consideration for the future positron source designs (ILC and CLIC projects). In this framework the first study was performed in the context of the SuperB project, where a capture efficiency increase close to one order of magnitude was obtained by using a large L-band section operated in higher order mode, to ensure a good longitudinal matching with the bunch length at the AMD exit~\cite{poirier2010positron}.
That would made the design, construction and integration of the higher-field DC solenoid around the capture system more challenging, requiring eventual application of the SC technology to be considered.

Based on the theoretical consideration, to capture a large phase space volume, the AMD with higher-field tapered solenoids around the target and downstream is desired. To reach higher capture efficiency, the target enclosed in a high-field solenoid magnet should be considered. 
Presently, the highest field of the matching device is obtained with the FC, very often in combination with the bridge coils. The peak magnetic field for such devices is typically about 5-6~T~\cite{ kulikov1991slc, Liu:2021iay}, while the ultimate performance is reached at BINP, where the FC used in the VEPP-5 injector complex can generate a 10~T peak field ~\cite{astrelina2008production,lapik2004some}. The contribution from the bridge coils is, usually, at the order of 1-2~T. 

As stated before, for positron sources, operating with typical $\mu$s pulses, FC are used as the matching devices. However, as an example, for the undulator-based positron source for the ILC, driven at 5~Hz with pulse duration of about 1~ms, a matching device compatible with such a regime is required. In general, the FC are machined from bulky Cu-conductors and this would, for very long pulses, lead to strongly time-varying current densities (time-varying skin effect), mainly in radial direction, and, thus, to time-varying magnetic field along the axis of the matching device. Using a single coil solenoid with a relatively thin Cu-conductor of about 1~cm$^2$ cross section would reach after a pulse rise time of about 2~ms a quasi constant current density distribution over the cross section of the conductor, similar to a DC case. Thus, trapezoidal pulses with a duration of about 4-5~ms with an intermediate flat top of 1~ms seem to be adequate. With the ILC repetition rate of 5~Hz, average power due to Ohmic heating of the coil will be tolerable. This power can conveniently be removed through a water cooling channel along the centre of the hollow Cu-conductor.
Recent tracking simulations of the positrons from the target-converter up to the DR with such tapered pulsed solenoid as the matching device report the positron yield of well above the nominal value $\geq$~1.5~$N_{e^+}/N_{e^-}$.
As the proposed pulsed solenoid is operated very close to the upstream, fast rotating target (at 100 m/s), issues of magnetically-induced pulsed braking forces and induced average power in the target have to be carefully considered. 
Studies of these effects are underway with appropriate finite element method simulation codes (COMSOL Multiphysics~\cite{comsol}) within the ILC Sources Working Group. 
Further investigations are planned in the future to address the current leads, strip lines, pulse forming network and the pulsed power supply. For this, one may profit from technology applied in the past for pulsed magnets, magnetic horns, pulsed Lithium lenses or similar devices.

Given the rate of progress in magnet R\&D, pushing the field of both solenoids (matching device and DC solenoid around the capture linac) beyond the state of the art, would improve considerably the positron source performance. New solutions to the high-field solenoids, thus, should be investigated. In particular, the innovative proposal to use a superconducting (SC) solenoid as the matching device for capture system is of great interest. The proposed technology has never been tested before in such a context and can allow reaching higher magnetic field on the target-converter and DC operation compared to the typically used FC. 
The SC solenoid based on High-Temperature Superconductor (HTS) technology is currently being explored in the framework of CHART (Swiss Accelerator Research and Technology) collaboration for the FCC-ee positorn source.
A feasibility and engineering study is ongoing, including considerations of the power distribution in target vicinity and radiation arising from beam interaction with the target. 
Depending on the obtained results, fully integrated design including the target, SC solenoid-based matching device and capture linac should be developed with further demonstration of the technology and performances.

Another important factor to increase the capture section efficiency, is the possibility to provide high solenoidal fields around the first accelerating structures beyond the 0.5~T, while the adiabatic damping acts, at higher energy, as a primary emittance damper. In this context, strong synergies can be seen with the Muon Collider Program (MAP) on the ionization cooling section, requiring high-gradient cavities immersed in high solenoidal fields ~\cite{PhysRevSTAB.8.072001, PhysRevAccelBeams.23.072001}.


Thanks to the fast-paced development of plasma wakefield acceleration technology, the application of the plasma lens to positron beam capture renewed its interest.
Although, the use of the plasma lens as the AMD has been recently explored theoretically~\cite{formela2021designing} being very promising, no big attention to technology problems has been yet dedicated.



\subsection{Beam dynamics and modelling}

As stated before, the positron beam has a large emittance and energy spread that is challenging to transport through the capture section, chicane, pre-accelerator and other stages including injection into the DR. 
The beam dynamics models should take into account the space charge effect between the particles and the wake field effect from the accelerating structures.
In case of  the multi-bunch operation, particular attention should be paid to longitudinal collective effect such as beam loading. Therefore, the appropriate  RF cavity design and effective beam loading compensation schemes should be considered. 

In the framework of the ILC undulator-based positron source, the significant positron bunch charge (several nC) results in high pulse beam loading for normal-conducting accelerating structures in positron pre-accelerator (in this definition it includes positron capture linac, pre-accelerator, beams separation and accelerator)~\cite{Paramonov2006}. In this context, the beam loading compensation method with the detuning angle has been developed to be applied for the positron capture linac of the ILC electron-driven positron source~\cite{Kuriki2019}.

To match the positron beam longitudinal phase space with the acceptance of the capture linac, it is efficient to use a design with the first RF structure in decelerating phase~\cite{aune1979new}. This technique should deserve more attention in its potentiality as demonstrated in the preliminary studies for the SuperB and CLIC projects and currently used at SuperKEKB positron capture section.

\paragraph{Positron source design}
In general, a modular approach is used in designing and modelling the positron sources. Most of the time, the various codes are employed for the individual sub-system (modules) of the positron sources (see Figure~\ref{fig:main_layout}). Being written in different languages and using the different standards for the data I/O make it difficult performing the start-to-end simulations and global optimization. In this context, some routines have been already developed and successfully applied~\cite{HAN201983, Zhao:2021tfi}.

The most widely used numerical tools to study the particle interaction with matter and, so the positron generation in the conventional positron production scheme are Geant4~\cite{ALLISON2016186} and FLUKA~\cite{osti_877507}. These codes are also used to simulate the energy deposition in the target and beamline components.
The standard tracking codes are used afterwards for positron capture simulations: ASTRA~\cite{astra}, RF-track~\cite{latina2016rf}, GPT~\cite{gpt}, Geant4, etc.. Many tracking tools are available for design and simulations of linear accelerators, which are, usually, used to simulate the positron beam pre-acceleration from a few hundreds of MeV and until injection to the DR: MAD-X~\cite{Deniau:ICAP12-THAAI3}, SAD~\cite{sad}, PLACET~\cite{latina2012improvements}, etc..
Owing to complexity of the positron source design, promising results have been also obtained by global parameter optimization with genetic algorithm developed in the code GIOTTO~\cite{Bacci:2016few}.

Some of the simulation tools are also incorporating the functionality to characterize the impact of irradiation on the materials via radiation load and  dpa analysis (available in FLUKA). In addition, the commercial package ANSYS~\cite{ansys} is used to asses the thermal load in the target and failure criteria in material by means of the thermo-mechanical stresses.

The modeling of the novel schemes for positron production requires additional simulation tools for photon generation.
The simulations of the crystal-based positron sources rely mainly on the in-house codes. Simulation models \cite{XAVIER1990278,PhysRevA.86.042903,PhysRevAccelBeams.22.064601} including the quasiclassical Baier-Katkov method \cite{Baier_Katkov_Strakhovenko} well validated experimentally \cite{PhysRevA.86.042903,PhysRevAccelBeams.22.064601,PhysRevLett.121.021603,CHEHAB200241} allow one to simulate charged particle dynamics and coherent effects of radiation and pair production in oriented crystals. These models are fundamental for the simulations of electromagnetic shower in the hybrid crystal-based positron source.
An effort to implement these models into the Geant4 simulation toolkit is currently ongoing. This will allow one to simulate an entire experimental setup of the crystal-based positron source. This effort will bring such application to a large scientific and industrial community and under a free Geant4 license.
For the undulator-based positron sources, Helical Undulator Synchrotron Radiation (HUSR) software~\cite{newton2010rapid} is used to simulate the photon energy distribution inside the undulator. It was developed by David Newton at the Cockcroft Institute (United Kingdom) and used for the ILC polarized positron source design. The other available widespread simulation code to calculate undulator radiation is SPECTRA~\cite{tanaka2001spectra} developed by Takashi Tanaka at the SPring-8.
The simulation code CAIN~\cite{Yokoya} developped by Kaoru Yokoya at KEK is usually used to simulate the Compton scattering process for the Compton scattering-based polarized positron sources. Recently, this option has been added in a tracking code RF-track~\cite{latina2016rf} developed by Andrea Latina at CERN.

A key aspect of the positron source design, thus, is the interdisciplinary simulations. The modeling of the positron sources involves several disciplines and requires, in general, a multiphysics approach.
A particle/matter interaction simulations coupled to beam dynamics codes for particle tracking in the electromagnetic fields together with the tools dedicated to calculation of the beam-induced damages in the target material are systematically in use. 

\paragraph{Advanced optimization techniques}
Another innovative aspect, that should be explored in this field is the Artificial Intelligence (AI) global optimization of the positron injector parameters including the electron drive beam and the final system acceptance. Recently, AI has already been applied for optimization, online control and tuning of modern particle accelerators, such as free electron lasers and storage ring light source~\cite{PhysRevLett.121.044801}. In terms of number of necessary evaluations of the optimisation function, one of the very efficient methods is Bayesian Optimisation (BO). Currently BO is broadly used in search of the optimal set of hyper parameters for so-called optimal bias-variance trade-off of Machine Learning (ML) model (huge amount of data, which slows done the ML training). BO was successfully applied to optimize the beam intensity at SwissFEL with up to 40 control parameters keeping safe operation constraints~\cite{20.500.11850/385955}. Proposed method was able to consistently outperform the typical baseline parameter sets. Another very recent example of  BO application  is in Laser-Plasma research  for tuning the “unknown parameters” of the new ionisation injection acceleration scheme both in  numerical simulation and in the corresponding experiment. Optimal conditions permitted generation of stable electron beam with subpercent level of beam energy spread~\cite{PhysRevLett.126.104801}. 



\section{Future directions of studies and R\&D}
\label{sec:fut_dir}
In order to achieve the required high luminosity at the future colliders, it is necessary to improve each single important aspect of the positron sources as the design: the modeling and the construction/operation of the innovative targets, the capture systems and the matching in the damping rings. 
Moreover, the positron source community should consolidate the effort and explore different methods of positron production, both with classical techniques and novel/exotic ones~\cite{chen1992coherent, PhysRevLett.79.1626, PhysRevLett.97.175003}. 
This should open the way primarily for future high-energy physics applications requiring orders of magnitude higher intensity, and for considering future hadronic applications (including the EIC) requiring both polarization and intensity. 
The request for the polarized positron beams results in an additional degree of complexity, which makes the polarized positron sources rather large and expensive facilities.

The different possible directions to face these challenges, and to cross the transition from the old concept to the novel positron sources, have been summarized in this article. 
Concerning the target, it is important to develop an important effort in modeling the fast regimes of the shock waves with a comprehensive understanding of their dynamics in the target and its holder. New target technologies have to be studied and developed, like the liquid- and the gas-jet targets. Target cooling is of fundamental importance and the effort in prototyping engineering solutions, such as the rotating wheel targets, is very important also if time and cost resources demanding. In the same framework, it is crucial to test the possibility to use SC magnets as matching devices and to increase the solenoidal fields around the capture linac accelerating sections. Primary beams should profit of the new technologies of plasma wakefield acceleration and high-gradient cavities to decrease costs and increase their performance. The realization of the toolkit, that is capable to perform the start-to-end positron source modeling and optimization remains to be developed. Up to now, the simulation tools are developed generally in-house and only for the particular applications/projects. New concepts and schemes for positron sources and polarised beams should be investigated when possible. 
All these challenges require a coordinated and collective effort in a multidisciplinary context focused on the development of appropriate simulation codes and realization of the different technology test stands, to prove the effective efficiency of the solutions and designs proposed.

A special task is the investigation of advanced accelerator concepts for the  positron beams relying on high-gradient plasma and laser accelerators.
Owing to high-luminosity demand, positron beams have to be generated and accelerated at a nC bunch charge with about 100~nm normalized transverse emittance and energy spread well below percent level. 
Although in the past few decades advances in the field are extremely rapid, this milestone remains to be achieved.





\acknowledgments

This work was supported, in part, by the European Union’s Horizon 2020 Research and Innovation programme under Grant Agreement No 101004730.
I.~C. wish to acknowledge financial support from ANR (Agence Nationale de la Recherche) under Grant No: ANR-21-CE31-0007.


\bibliographystyle{unsrt}
\bibliography{bibliography}

\begin{thebibliography}{100}

\bibitem{europeanstrategygroup2020physics}
European~Strategy for Particle Physics Preparatory~Group.
\newblock Physics briefing book, 2020.

\bibitem{adolphsen2022european}
C.~Adolphsen, D.~Angal-Kalinin, T.~Arndt, et~al.
\newblock European strategy for particle physics -- accelerator {R\&D} roadmap.
\newblock {\em arXiv:2201.07895}, 2022.

\bibitem{Bartosik_2020}
N.~Bartosik, A.~Bertolin, L.~Buonincontri, M.~Casarsa, F.~Collamati,
  A.~Ferrari, A.~Ferrari, A.~Gianelle, D.~Lucchesi, N.~Mokhov, M.~Palmer,
  N.~Pastrone, P.~Sala, L.~Sestini, and S.~Striganov.
\newblock Detector and physics performance at a muon collider.
\newblock {\em Journal of Instrumentation}, 15(05):P05001--P05001, may 2020.

\bibitem{Peach}
Ken Peach.
\newblock {The Future of Particle Physics: The LHC and Beyond}.
\newblock pages 625--630. 6 p, 2020.

\bibitem{FCC:2018byv}
A.~Abada et~al.
\newblock {FCC Physics Opportunities}: {Future Circular Collider Conceptual
  Design Report Volume 1}.
\newblock {\em Eur. Phys. J. C}, 79(6):474, 2019.

\bibitem{clic_phys}
Eva Sicking and Rickard Str\"om.
\newblock {From precision physics to the energy frontier with the Compact
  Linear Collider}.
\newblock {\em Nature Phys.}, 16(4):386--392, 2020.

\bibitem{BAUER1990300}
W.~Bauer, J.~Briggmann, H.-D. Carstanjen, S.~Connell, W.~Decker, J.~Diehl,
  K.~Maier, J.~Major, H.-E. Schaefer, A.~Seeger, H.~Stoll, and E.~Widmann.
\newblock The {Stuttgart} positron beam, its performance and recent
  experiments.
\newblock {\em Nuclear Instruments and Methods in Physics Research Section B:
  Beam Interactions with Materials and Atoms}, 50(1):300--306, 1990.

\bibitem{golge2012review}
S~Golge, B~Vlahovic, et~al.
\newblock Review of low-energy positron beam facilities.
\newblock {\em Proc. of IPAC}, 12:1464--1466, 2012.

\bibitem{cassidy2006accumulator}
DB~Cassidy, SHM Deng, RG~Greaves, and AP~Mills~Jr.
\newblock Accumulator for the production of intense positron pulses.
\newblock {\em Review of Scientific Instruments}, 77(7):073106, 2006.

\bibitem{CHARLTON2021164657}
M.~Charlton, J.J. Choi, M.~Chung, P.~Cladé, et~al.
\newblock Positron production using a 9 mev electron linac for the gbar
  experiment.
\newblock {\em Nuclear Instruments and Methods in Physics Research Section A:
  Accelerators, Spectrometers, Detectors and Associated Equipment}, 985:164657,
  2021.

\bibitem{hessami:napac2019-moplh23}
R.M. Hessami and S.J. Gessner.
\newblock {An Analysis of Potential Compact Positron Beam Source}.
\newblock In {\em Proc. NAPAC'19}, number~4 in North American Particle
  Accelerator Conference, pages 220--223. JACoW Publishing, Geneva,
  Switzerland, 10 2019.
\newblock https://doi.org/10.18429/JACoW-NAPAC2019-MOPLH23.

\bibitem{PhysRevLett.90.214801}
B.~E. Blue, C.~E. Clayton, C.~L. O'Connell, F.-J. Decker, M.~J. Hogan,
  C.~Huang, R.~Iverson, C.~Joshi, T.~C. Katsouleas, W.~Lu, K.~A. Marsh, W.~B.
  Mori, P.~Muggli, R.~Siemann, and D.~Walz.
\newblock Plasma-wakefield acceleration of an intense positron beam.
\newblock {\em Phys. Rev. Lett.}, 90:214801, May 2003.

\bibitem{PhysRevLett.90.205002}
M.~J. Hogan, C.~E. Clayton, C.~Huang, P.~Muggli, S.~Wang, B.~E. Blue, D.~Walz,
  K.~A. Marsh, C.~L. O'Connell, S.~Lee, R.~Iverson, F.-J. Decker, P.~Raimondi,
  W.~B. Mori, T.~C. Katsouleas, C.~Joshi, and R.~H. Siemann.
\newblock Ultrarelativistic-positron-beam transport through meter-scale
  plasmas.
\newblock {\em Phys. Rev. Lett.}, 90:205002, May 2003.

\bibitem{PhysRevLett.101.055001}
P.~Muggli, B.~E. Blue, C.~E. Clayton, F.~J. Decker, M.~J. Hogan, C.~Huang,
  C.~Joshi, T.~C. Katsouleas, W.~Lu, W.~B. Mori, C.~L. O'Connell, R.~H.
  Siemann, D.~Walz, and M.~Zhou.
\newblock Halo formation and emittance growth of positron beams in plasmas.
\newblock {\em Phys. Rev. Lett.}, 101:055001, Jul 2008.

\bibitem{doche2017acceleration}
A~Doche, C~Beekman, S{\'e}bastien Corde, JM~Allen, CI~Clarke, J~Frederico,
  SJ~Gessner, SZ~Green, MJ~Hogan, B~O’Shea, et~al.
\newblock Acceleration of a trailing positron bunch in a plasma wakefield
  accelerator.
\newblock {\em Scientific reports}, 7(1):1--7, 2017.

\bibitem{corde2015multi}
S{\'e}bastien Corde, E~Adli, JM~Allen, W~An, CI~Clarke, CE~Clayton,
  JP~Delahaye, J~Frederico, S~Gessner, SZ~Green, et~al.
\newblock Multi-gigaelectronvolt acceleration of positrons in a self-loaded
  plasma wakefield.
\newblock {\em Nature}, 524(7566):442--445, 2015.

\bibitem{gessner2016demonstration}
Spencer Gessner, Erik Adli, James~M Allen, Weiming An, Christine~I Clarke,
  Chris~E Clayton, Sebastien Corde, JP~Delahaye, Joel Frederico, Selina~Z
  Green, et~al.
\newblock Demonstration of a positron beam-driven hollow channel plasma
  wakefield accelerator.
\newblock {\em Nature communications}, 7(1):1--6, 2016.

\bibitem{PhysRevResearch.3.043063}
C.~S. Hue, G.~J. Cao, I.~A. Andriyash, A.~Knetsch, M.~J. Hogan, E.~Adli,
  S.~Gessner, and S.~Corde.
\newblock Efficiency and beam quality for positron acceleration in loaded
  plasma wakefields.
\newblock {\em Phys. Rev. Research}, 3:043063, Oct 2021.

\bibitem{alejo2019laser}
Aaron Alejo, Roman Walczak, and Gianluca Sarri.
\newblock Laser-driven high-quality positron sources as possible injectors for
  plasma-based accelerators.
\newblock {\em Scientific reports}, 9(1):1--10, 2019.

\bibitem{Bharadwaj2001slc}
V.K. Bharadwaj, Y.K. Batygin, J.C. Sheppard, D.C. Schultz, S.~Bodenstein,
  J.~Gallegos, R.~Gonzales, J.~Ledbetter, M.~Lopez, R.~Romero, T.~Romero,
  R.~Rutherford, and S.~Maloy.
\newblock Analysis of beam-induced damage to the {SLC} positron production
  target.
\newblock In {\em Proceedings of the 2001 Particle Accelerator Conference},
  volume~3, pages 2123--2125, 2001.

\bibitem{helm1962adiabatic}
R.H. Helm.
\newblock Adiabatic approximation for dynamics of a particle in the field of a
  tapered solenoid.
\newblock Technical report, Stanford Univ., Calif. Stanford Linear Accelerator
  Center, 1962.

\bibitem{chehab1978second}
R~Chehab.
\newblock A second order calculation of the adiabatic invariant of a charged
  particle spiraling in a longitudinal magnetic field.
\newblock {\em Journal of Mathematical Physics}, 19(5):937--941, 1978.

\bibitem{Chehab:197428}
R~Chehab.
\newblock {Positron sources}.
\newblock {\em CAS - CERN Accelerator School: 3rd General Accelerator Physics
  Course}, page 30 p, Jan 1989.

\bibitem{ADA}
J.~Haissinski.
\newblock {\em Expériences sur l’anneau de collisions AdA}.
\newblock Phd, Physique des accélérateurs [physics.acc-ph]., 1965.

\bibitem{ACO}
P.~C. Marin.
\newblock Storage rings at {Orsay}.
\newblock {\em Nuclear Instruments and Methods in Physics Research Section A:
  Accelerators, Spectrometers, Detectors and Associated Equipment},
  266(19072981):18--23, 1988.

\bibitem{DCI}
J.~Le~Duff, M.~P. Level, P.~Marin, M.~Sommer, and H.~Zyngier.
\newblock {Status Report on {D.C.I.}}
\newblock In {\em {10th International Conference on High-Energy Accelerators}},
  pages 469--477, 1977.

\bibitem{SPEAR}
J.M. Paterson et~al.
\newblock {{SPEAR II} performance}.
\newblock In {\em {IEEE (Inst Electr. Electron. Eng.) Trans Nucl Sci,
  NS-22(3)}}, pages 1366--1369, 1975.

\bibitem{ADONE}
F.~Amman, R.~Andreani, et~al.
\newblock {ADONE}-the {F}rascati 1.5 {GeV} electron positron storage ring. note
  no. 285.
\newblock Technical Report LNF-65/26, Laboratori Nazionali, Frascati, Italy,
  1965.

\bibitem{VEPP}
VEPP-4 Group.
\newblock {}.
\newblock In {\em {11th International Conference on High-Energy Accelerators:
  Geneva, Switzerland}}, pages 38--42, 1980.

\bibitem{LEP_90}
C.~Bourat, H.~Braun, and L.~Rinolfi.
\newblock New optics of the {LEP} injector linac for positron production.
\newblock In {\em Proc. 4th European Particle Accelerator Conf. (EPAC'94)},
  pages 704--707. JACoW Publishing, Jun.-Jul. 1994.

\bibitem{Oide:2009zz}
K.~Oide.
\newblock {{KEKB} {B}-factory, the luminosity frontier}.
\newblock {\em Prog. Theor. Phys.}, 122:69--80, 2009.

\bibitem{AKAI2018188}
Kazunori Akai, Kazuro Furukawa, and Haruyo Koiso.
\newblock Superkekb collider.
\newblock {\em Nuclear Instruments and Methods in Physics Research Section A:
  Accelerators, Spectrometers, Detectors and Associated Equipment},
  907:188--199, 2018.
\newblock Advances in Instrumentation and Experimental Methods (Special Issue
  in Honour of Kai Siegbahn).

\bibitem{PEP-II}
A.~Hutton and M.S. Zisman.
\newblock {PEP-II}: an asymmetric {B} factory based on {PEP}.
\newblock In {\em Proceedings of the 1991 IEEE Particle Accelerator
  Conference}, volume~1, pages 84--86, 1991.

\bibitem{PEP-pos}
Eric Bloom, F.~Bulos, T.~Fieguth, Gary Godfrey, G.~Loew, and R.~Miller.
\newblock Progress on {PEP-II} injection {R\&D}.
\newblock In {\em Proceedings of the 1993 IEEE Particle Accelerator
  Conference}, volume~4, pages 3084--3086, 06 1993.

\bibitem{Clendenin:1988np}
J.~E. Clendenin et~al.
\newblock {SLC} positron source startup.
\newblock In {\em {14th International Linear Accelerator Conference}}, 9 1988.

\bibitem{VARIOLA201421}
A.~Variola.
\newblock Advanced positron sources.
\newblock {\em Nuclear Instruments and Methods in Physics Research Section A:
  Accelerators, Spectrometers, Detectors and Associated Equipment}, 740:21--26,
  2014.
\newblock Proceedings of the first European Advanced Accelerator Concepts
  Workshop 2013.

\bibitem{slc-target}
{Reuter E., Mansour D., Porter T., Sax W., and Szumillo A.}
\newblock {Mechanical Design and Development of a High Power Target System for
  the SLC Positron Source }.
\newblock SLAC-PUB--5369.

\bibitem{clendenin1996compendium}
J~Clendenin, L~Rinolfi, K~Takata, and DJ~Warner.
\newblock Compendium of scientific linacs.
\newblock {\em CERN/PS}, 96:32, 1996.

\bibitem{prcom1}
Y.~Enomoto.
\newblock {Private Communication}, 2022.

\bibitem{dafne-linac}
Fernando Sannibale, M.~Vescovi, R.~Boni, Fiorella Marcellini, and G.~Vignola.
\newblock Dafne linac commissioning results.
\newblock 02 2022.

\bibitem{dafne-linac2}
Robert Boni, Fabio Marcellini, F~Sannibale, and Michele Vescovi.
\newblock Dafne linac operational performance.
\newblock 1998.

\bibitem{bepc-pos}
Pei Guoxi.
\newblock Progress of the {BEPCII} linac upgrade.
\newblock In {\em Proceedings of the 2005 Particle Accelerator Conference},
  pages 2416--2418, 2005.

\bibitem{LIL}
Rudolf Bossart, Delahaye jean pierre, J~Godot, J~Madsen, Peter Pearce, A~Riche,
  and Louis Rinolfi.
\newblock The lep injector linac.
\newblock 01 1990.

\bibitem{LIL-robert}
Robert Chehab.
\newblock {Quelques specification concernant la source de positrons du LIL}.
\newblock Technical Report LAL/PI/55-82/T, LAL, Orsay, 1982.

\bibitem{astrelina2008production}
KV~Astrelina, MF~Blinov, TA~Vsevolozhskaya, NS~Dikanskii, FA~Emanov, RM~Lapik,
  PV~Logachev, PV~Martyshkin, AV~Petrenko, TV~Rybitskaya, et~al.
\newblock Production of intense positron beams at the vepp-5 injection complex.
\newblock {\em Journal of Experimental and Theoretical Physics}, 106(1):77--93,
  2008.

\bibitem{prcom2}
R.~Chehab.
\newblock {Private Communication}, 2022.

\bibitem{toge1995kek}
Nobu Toge.
\newblock {KEK B-Factory} design report.
\newblock Technical Report KEK Report 95-7, National Laboratory for High Energy
  Physics (KEK), 1995.

\bibitem{satoh2016commissioning}
M~Satoh, M~Akemoto, Y~Arakida, D~Arakawa, N~Iida, H~Iwase, A~Enomoto,
  Y~Enomoto, S~Fukuda, Y~Funakoshi, et~al.
\newblock Commissioning status of {SuperKEKB} injector linac.
\newblock {\em Energy}, 100(2100):50, 2016.

\bibitem{kamitani2014superkekb}
Takuya Kamitani, Mitsuo Akemoto, Dai Arakawa, Yoshio Arakida, Atsushi Enomoto,
  Shigeki Fukuda, Yoshihiro Funakoshi, Kazuro Furukawa, Toshiyasu Higo,
  Hiroyuki Honma, et~al.
\newblock {SuperKEKB} positron source construction status.
\newblock In {\em 5th Int. Particle Accelerator Conf.(IPAC'14), Dresden,
  Germany, June 15-20, 2014}, pages 579--581. JACOW Publishing, Geneva,
  Switzerland, 2014.

\bibitem{kulikov1991slc}
AV~Kulikov, SD~Ecklund, and EM~Reuter.
\newblock {SLC} positron source pulsed flux concentrator.
\newblock Technical report, Stanford Linear Accelerator Center, 1991.

\bibitem{enomoto:ipac2021-wepab144}
Y.~Enomoto, K.~Abe, N.~Okada, and T.~Takatomi.
\newblock {A New Flux Concentrator Made of {Cu} Alloy for the {SuperKEKB}
  Positron Source}.
\newblock In {\em Proc. IPAC'21}, number~12 in International Particle
  Accelerator Conference, pages 2954--2956. JACoW Publishing, Geneva,
  Switzerland, 08 2021.
\newblock https://doi.org/10.18429/JACoW-IPAC2021-WEPAB144.

\bibitem{suwada2021first}
Tsuyoshi Suwada, Muhammad~Abdul Rehman, and Fusashi Miyahara.
\newblock First simultaneous detection of electron and positron bunches at the
  positron capture section of the {SuperKEKB} factory.
\newblock {\em Scientific Reports}, 11(1):1--9, 2021.

\bibitem{Shiltsev2021}
V.~Shiltsev and F.~Zimmermann.
\newblock Modern and future colliders.
\newblock {\em Reviews of Modern Physics}, 93(1), Mar 2021.

\bibitem{LHCE2012}
J~L Abelleira~Fernandez, C~Adolphsen, A~N Akay, H~Aksakal, J~L Albacete,
  S~Alekhin, P~Allport, V~Andreev, R~B Appleby, E~Arikan, and et~al.
\newblock A {Large Hadron Electron Collider at CERN} report on the physics and
  design concepts for machine and detector.
\newblock {\em Journal of Physics G: Nuclear and Particle Physics},
  39(7):075001, Jul 2012.

\bibitem{moortgat2008polarized}
Gudrid Moortgat-Pick, T~Abe, G~Alexander, B~Ananthanarayan, AA~Babich,
  V~Bharadwaj, D~Barber, A~Bartl, A~Brachmann, Si~Chen, et~al.
\newblock Polarized positrons and electrons at the linear collider.
\newblock {\em Physics Reports}, 460(4-5):131--243, 2008.

\bibitem{grames2018polarized}
{\em {Proceedings, International Workshop on Physics with Positrons at
  Jefferson Lab (JPos17)}: {Newport News, VA, USA, September 12-15, 2017}}, 5
  2018.

\bibitem{PhysRevLett.116.214801}
D.~Abbott, P.~Adderley, A.~Adeyemi, et~al.
\newblock Production of highly polarized positrons using polarized electrons at
  {MeV} energies.
\newblock {\em Phys. Rev. Lett.}, 116:214801, May 2016.

\bibitem{Nature}
M.~Benedikt, A.~Blondel, P.~Janot, and F.~Zimmermann.
\newblock {Future Circular Colliders succeeding the LHC}.
\newblock {\em Nat. Phys.}, 16:402--407, 2020.

\bibitem{FCC_CDR_2}
A.~Abada et~al.
\newblock {FCC}-ee: The lepton collider.
\newblock {\em The European Physical Journal Special Topics}, 228(2):261--623,
  2019.

\bibitem{alesini2019positron}
D~Alesini, M~Antonelli, ME~Biagini, M~Boscolo, OR~Blanco-Garc{\'\i}a, A~Ciarma,
  R~Cimino, M~Iafrati, A~Giribono, S~Guiducci, et~al.
\newblock Positron driven muon source for a muon collider.
\newblock {\em arXiv preprint arXiv:1905.05747}, 2019.

\bibitem{Zhao:2735292}
Yongke Zhao, Andrea Latina, Steffen Doebert, Daniel Schulte, and Lianliang Ma.
\newblock {Optimisation of the {CLIC} positron source at the 1.5 {TeV} and 3
  {TeV} stages}.
\newblock Technical report, CERN, Geneva, Sep 2020.

\bibitem{riemann2020updated}
Sabine Riemann, Peter Sievers, Gudrid Moortgat-Pick, and Andriy Ushakov.
\newblock Updated status of the undulator-based {ILC} positron source.
\newblock {\em arXiv:2002.10919}, 2020.

\bibitem{resultsILCconv}
{T. Omori}.
\newblock {Overview of R\&D status of ILC E-Driven positron source}.
\newblock International Workshop on Future Linear Colliders, LCWS2021.

\bibitem{zimmermann:ipac12-weppr076}
F.~Zimmermann et~al.
\newblock {Positron Options for the Linac-ring LHeC}.
\newblock In {\em Proc. IPAC'12}, pages 3108--3110. JACoW Publishing, Geneva,
  Switzerland.

\bibitem{Alesini:2019tlf}
D.~Alesini et~al.
\newblock {Positron driven muon source for a muon collider}.
\newblock {\em arXiv:1905.05747}, 5 2019.

\bibitem{Wang:2017mjk}
Dou Wang, Yunlong Chi, Jie Gao, Xiaoping Li, Cai Meng, Guoxi Pei, and Jingru
  Zhang.
\newblock {Design Study on CEPC Positron Damping Ring and Bunch Compressor}.
\newblock In {\em {8th International Particle Accelerator Conference}}, 5 2017.

\bibitem{balakinconversion}
V.E. Balakin and A.A. Mikhailichenko.
\newblock Conversion system for obtaining highly polarized positrons and
  electrons at high energy.
\newblock Technical report, Budker INP preprint 79-85, Novosibirsk, 1979.

\bibitem{Balakin:1988se}
V.~E. Balakin and A.~A. Mikhailichenko.
\newblock { VLEPP: The conversion System for Obtaining Highly Polarized
  Electrons and Positrons}.
\newblock In {\em Proceedings of the 12th International Conference on
  High-Energy Accelerators}, pages 127--131, 1983.
\newblock Prepared for 12th International Conference on High-Energy
  Accelerators, Batavia, Illinois, 11-16 Aug 1983.

\bibitem{mikhailichenko1925conversion}
A.A. Mikhailichenko.
\newblock {\em Conversion System for Obtaining Polarized Electrons and
  Positrons at High Energy Translation}.
\newblock PhD thesis, Acasemy of science USSR, Budker Institute of Nuclear
  Physics, Novosibirsk, 1986 (translated 2002).

\bibitem{ILC2013TDRvol3Accelerator}
Chris Adolphsen, Maura Barone, Barry Barish, et~al.
\newblock The {International Linear Collider Technical Design Report} - volume
  3.{II}: Accelerator baseline design, 2013.

\bibitem{Bambade2019ilc}
Philip Bambade, Tim Barklow, Ties Behnke, et~al.
\newblock The {International Linear Collider}: A global project.
\newblock {\em arXiv:1903.01629}, 2019.

\bibitem{alexander2008observation}
G.~Alexander et~al.
\newblock Observation of polarized positrons from an undulator-based source.
\newblock {\em Physical review letters}, 100(21):210801, 2008.

\bibitem{Alexander:2009nb}
G.~Alexander et~al.
\newblock {Undulator-Based Production of Polarized Positrons}.
\newblock {\em Nucl. Instrum. Meth.}, A610:451--487, 2009.

\bibitem{Scott2011Undulator}
D.~J. Scott, J.~A. Clarke, D.~E. Baynham, V.~Bayliss, T.~Bradshaw, G.~Burton,
  A.~Brummitt, S.~Carr, A.~Lintern, J.~Rochford, O.~Taylor, and
  Y.~Ivanyushenkov.
\newblock Demonstration of a high-field short-period superconducting helical
  undulator suitable for future {TeV}-scale linear collider positron sources.
\newblock {\em Phys. Rev. Lett.}, 107:174803, Oct 2011.

\bibitem{Alharbi2021ipac}
K~Alharbi et~al.
\newblock Design of photon masks for the {ILC} positron source.
\newblock {\em Proceedings of IPAC2021, Campinas, Brazil}, pages 3834--3836,
  2021.

\bibitem{Nagoshi2020nima}
H.~Nagoshi, M.~Kuribayashi, M.~Kuriki, P.V. Martyshkin, T.~Omori, T.~Takahashi,
  M.~Yamakata, and K.~Yokoya.
\newblock A design of an electron driven positron source for the {International
  Linear Collider}.
\newblock {\em Nuclear Instruments and Methods in Physics Research Section A:
  Accelerators, Spectrometers, Detectors and Associated Equipment}, 953:163134,
  2020.

\bibitem{bessonov1996method}
E.G. Bessonov and A.A. Mikhailichenko.
\newblock A method of polarized positron beam production.
\newblock In {\em Proceedings of the 5th European Particle Accelerator
  Conference}, page 1516, 1996.

\bibitem{okugi1996proposed}
T.~Okugi, Y.~Kurihara, M.~Chiba, A.~Endo, R.~Hamatsu, T.~Hirose, T.~Kumita,
  T.~Omori, Y.~Takeuchi, and M.~Yoshioka.
\newblock Proposed method to produce a highly polarized e+ beam for future
  linear colliders.
\newblock {\em Japanese journal of applied physics}, 35(part 1):3677--3680,
  1996.

\bibitem{PhysRevLett.96.114801}
T.~Omori, M.~Fukuda, T.~Hirose, Y.~Kurihara, R.~Kuroda, M.~Nomura, A.~Ohashi,
  T.~Okugi, K.~Sakaue, T.~Saito, J.~Urakawa, M.~Washio, and I.~Yamazaki.
\newblock Efficient propagation of polarization from laser photons to positrons
  through compton scattering and electron-positron pair creation.
\newblock {\em Phys. Rev. Lett.}, 96:114801, Mar 2006.

\bibitem{rinolfi2009clic}
L.~Rinolfi, H.H. Braun, Y.~Papaphilippou, D.~Schulte, A.~Vivoli, F.~Zimmermann,
  F.~Antoniou, I.R. Bailey, L.~Zang, E.V. Bulyak, et~al.
\newblock The clic positron sources based on compton schemes.
\newblock In {\em Proceedings of the 23rd Particle Accelerator Conference,
  Vancouver, Canada}. IEEE, 2009.

\bibitem{chaikovska2012polarized}
Iryna Chaikovska.
\newblock {\em Polarized positron sources for the future linear colliders}.
\newblock PhD thesis, Universit{\'e} Paris Sud-Paris XI, 2012.

\bibitem{zimmermann2009stacking}
F~Zimmermann, Y~Papaphilippou, L~Rinolfi, F~Antoniou, R~Chehab, M~Kuriki,
  T~Omori, J~Urakawa, A~Variola, A~Vivoli, et~al.
\newblock Stacking simulations for compton positron sources of future linear
  colliders.
\newblock In {\em Particle Accelerator Conference (PAC09)}, page MO6RFP064.
  Triumf Vancouver, 2009.

\bibitem{2019gamma}
W.~Płaczek, A.~Abramov, S.E. Alden, R.~Alemany~Fernandez, P.S. Antsiferov,
  A.~Apyan, H.~Bartosik, E.G. Bessonov, N.~Biancacci, J.~Bieroń, and et~al.
\newblock Gamma factory at {CERN} --- novel research tools made of light.
\newblock {\em Acta Physica Polonica B}, 50(6):1191, 2019.

\bibitem{Krasny:2690736}
M~W Krasny, A~Abramov, et~al.
\newblock {Gamma Factory Proof-of-Principle Experiment}.
\newblock Technical report, CERN, Geneva, Sep 2019.

\bibitem{chaikovska2016high}
I~Chaikovska, K~Cassou, R~Chiche, R~Cizeron, P~Cornebise, N~Delerue, D~Jehanno,
  F~Labaye, R~Marie, A~Martens, et~al.
\newblock High flux circularly polarized gamma beam factory: coupling a
  {Fabry-Perot} optical cavity with an electron storage ring.
\newblock {\em Scientific reports}, 6(1):1--9, 2016.

\bibitem{PhysRevAccelBeams22093501}
Cheikh~Fall Ndiaye, Kevin Cassou, et~al.
\newblock Low power commissioning of an innovative laser beam circulator for
  inverse compton scattering $\ensuremath{\gamma}$-ray source.
\newblock {\em Phys. Rev. Accel. Beams}, 22:093501, Sep 2019.

\bibitem{PhysRevAccelBeams.23.051301}
N.~Sudar, P.~Musumeci, A.~Ovodenko, A.~Murokh, M.~Polyanskiy, I.~Pogorelsky,
  M.~Fedurin, C.~Swinson, K.~Kusche, M.~Babzien, and M.~Palmer.
\newblock Burst mode {MHz} repetition rate inverse free electron laser
  acceleration.
\newblock {\em Phys. Rev. Accel. Beams}, 23:051301, May 2020.

\bibitem{Amoudry:20}
Lo\"{i}c Amoudry, Huan Wang, Kevin Cassou, Ronic Chiche, Kevin Dupraz,
  Aur\'{e}lien Martens, Daniele Nutarelli, Viktor Soskov, and Fabian Zomer.
\newblock Modal instability suppression in a high-average-power and
  high-finesse fabry--perot cavity.
\newblock {\em Appl. Opt.}, 59(1):116--121, Jan 2020.

\bibitem{PhysRevAccelBeams21121601B}
Pierre Favier, Lo\"{\i}c Amoudry, Kevin Cassou, et~al.
\newblock Optimization of a {Fabry-Perot} cavity operated in burst mode for
  compton scattering experiments.
\newblock {\em Phys. Rev. Accel. Beams}, 21:121601, Dec 2018.

\bibitem{rbtoptical}
A~Murokh RBT, P~Musumeci, JB~Rosenzweig, F~Zomer, K~Cassou, A~Martens,
  D~Nutarelli, and K~Dupraz LAL.
\newblock Optical energy recovery for a high duty cycle gamma ray source.
\newblock {\em Snowmass 2021 Letter of Interest (LOI) – AF6/AF4}, 2021.

\bibitem{CLICcdr}
M~Aicheler, P~Burrows, M~Draper, T~Garvey, P~Lebrun, K~Peach, N~Phinney,
  H~Schmickler, D~Schulte, and N~Toge.
\newblock {\em {A Multi-TeV Linear Collider Based on CLIC Technology: CLIC
  Conceptual Design Report}}.
\newblock CERN Yellow Reports: Monographs. CERN, Geneva, 2012.

\bibitem{katkov1998electromagnetic}
V.~N. Baier, V.~M. Katkov, and V.~M. Strakhovenko.
\newblock {\em Electromagnetic processes at high energies in oriented single
  crystals}.
\newblock World Scientific, 1998.

\bibitem{chehab1989study}
R~Chehab, F~Couchot, AR~Nyaiesh, F~Richard, and X~Artru.
\newblock Study of a positron source generated by photons from
  ultrarelativistic channeled particles.
\newblock In {\em Proceedings of the 1989 IEEE Particle Accelerator
  Conference,.'Accelerator Science and Technology}, pages 283--285. IEEE, 1989.

\bibitem{CHEHAB200241}
R~Chehab, R~Cizeron, C~Sylvia, V~Baier, et~al.
\newblock Experimental study of a crystal positron source.
\newblock {\em Physics Letters B}, 525(1):41--48, 2002.

\bibitem{ARTRU2005762}
X.~Artru, V.~Baier, K.~Beloborodov, et~al.
\newblock Summary of experimental studies, at {CERN}, on a positron source
  using crystal effects.
\newblock {\em Nuclear Instruments and Methods in Physics Research Section B:
  Beam Interactions with Materials and Atoms}, 240(3):762--776, 2005.

\bibitem{artru1996axial}
X~Artru, VN~Baier, TV~Baier, R~Chehab, M~Chevallier, E~Hourani, A~Jejcic,
  V~Katkov, R~Kirsch, K~Maier, et~al.
\newblock Axial channeling of relativistic electrons in crystals as a source
  for positron production.
\newblock {\em Nuclear Instruments and Methods in Physics Research Section B:
  Beam Interactions with Materials and Atoms}, 119(1-2):246--252, 1996.

\bibitem{PhysRevE.67.016502}
T.~Suwada, S.~Anami, R.~Chehab, et~al.
\newblock Measurement of positron production efficiency from a tungsten
  monocrystalline target using 4- and 8-{GeV} electrons.
\newblock {\em Phys. Rev. E}, 67:016502, Jan 2003.

\bibitem{ARTRU20083868}
X.~Artru, R.~Chehab, M.~Chevallier, V.M. Strakhovenko, A.~Variola, and
  A.~Vivoli.
\newblock Polarized and unpolarized positron sources for electron–positron
  colliders.
\newblock {\em Nuclear Instruments and Methods in Physics Research Section B:
  Beam Interactions with Materials and Atoms}, 266(17):3868--3875, 2008.
\newblock Radiation from Relativistic Electrons in Periodic Structures.

\bibitem{cheng2012positron}
Xu~Cheng-Hai, Robert Chehab, Peter Sievers, Xavier Artru, Michel Chevallier,
  Olivier Dadoun, Pei Guo-Xi, Vladimir~M Strakhovenko, and Alessandro Variola.
\newblock A positron source using an axially oriented crystal associated to a
  granular amorphous converter.
\newblock {\em Chinese Physics C}, 36(9):871, 2012.

\bibitem{Chaikovska:2019ztn}
Iryna Chaikovska et~al.
\newblock {Positron source for FCC-ee}.
\newblock In {\em {10th International Particle Accelerator Conference}}, page
  MOPMP003, 2019.

\bibitem{Afanasev2019physics}
A.~Afanasev et~al.
\newblock Physics with positron beams at {Jefferson Lab 12 GeV}.
\newblock {\em arXiv:1906.09419}, 2019.

\bibitem{Arrington2021physics}
J.~Arrington, M.~Battaglieri, A.~Boehnlein, S.~A. Bogacz, W.~K. Brooks,
  E.~Chudakov, I.~Cloet, R.~Ent, H.~Gao, J.~Grames, L.~Harwood, X.~Ji,
  C.~Keppel, G.~Krafft, R.~D. McKeown, J.~Napolitano, J.~W. Qiu, P.~Rossi,
  M.~Schram, S.~Stepanyan, J.~Stevens, A.~P. Szczepaniak, N.~Toro, and
  X.~Zheng.
\newblock Physics with {CEBAF} at 12 {GeV} and future opportunities.
\newblock {\em arXiv:2112.00060}, 2021.

\bibitem{Bertarelli:2016obw}
A.~Bertarelli.
\newblock {Beam-Induced Damage Mechanisms and their Calculation}.
\newblock In {\em {2014 Joint International Accelerator School}: {Beam Loss and
  Accelerator Protection}}, pages 159--227, 2016.

\bibitem{Mokhov:2016fdl}
N.~V. Mokhov and F.~Cerutti.
\newblock {Beam-Material Interaction}.
\newblock In {\em {2014 Joint International Accelerator School}: {Beam Loss and
  Accelerator Protection}}, pages 83--110, 2016.

\bibitem{cesarini2021theoretical}
Gianmario Cesarini, Mario Antonelli, Fabio Anulli, Matteo Bauce, Maria~Enrica
  Biagini, Oscar~R Blanco-Garc{\'\i}a, Manuela Boscolo, Fausto Casaburo,
  Gianluca Cavoto, Andrea Ciarma, et~al.
\newblock Theoretical modeling for the thermal stability of solid targets in a
  positron-driven muon collider.
\newblock {\em International Journal of Thermophysics}, 42(12):1--27, 2021.

\bibitem{maloy2001slc}
S~Maloy et~al.
\newblock {SLC} target analysis.
\newblock {\em LANL LA UR-01-1913}, 72, 2001.

\bibitem{stein2001thermal}
Werner Stein, A~Sunwoo, VK~Bharadwaj, DC~Schultz, and JC~Sheppard.
\newblock Thermal shock structural analyses of a positron target.
\newblock In {\em PACS2001. Proceedings of the 2001 Particle Accelerator
  Conference (Cat. No. 01CH37268)}, volume~3, pages 2111--2113. IEEE, 2001.

\bibitem{sievers2003stationary}
Peter Sievers.
\newblock A stationary target for the cern-neutrino-factory.
\newblock {\em Nuclear Instruments and Methods in Physics Research Section A:
  Accelerators, Spectrometers, Detectors and Associated Equipment},
  503(1-2):344--347, 2003.

\bibitem{Artru:359257}
X~Artru, R~Kirsch, R~Chehab, B~Johnson, P~Keppler, J~V Major, Louis Rinolfi,
  and A~Jejcic.
\newblock {Radiation-damage study of a monocrystalline tungsten positron
  converter}.
\newblock Technical Report CERN-PS-98-017-LP ; CLIC-Note-369 ; LAL-RT-98-02,
  Jun 1998.

\bibitem{bulos1985design}
F.~Bulos, H.~DeStaebler, S.~Ecklund, R.~Helm, H.~Hoag, H.~Le~Boutet, HL~Lynch,
  R.~Miller, KC~Moffeit, and S.L.A. Center.
\newblock Design of a high yield position source.
\newblock {\em Nuclear Science, IEEE Transactions on}, 32(5):1832--1834, 1985.

\bibitem{chehab1983adiabatic}
R.~Chehab, G.~Le~Meur, B.~Mouton, and M.~Renard.
\newblock An adiabatic matching device for the orsay linear positron
  accelerator.
\newblock {\em Nuclear Science, IEEE Transactions on}, 30(4):2850--2852, 1983.

\bibitem{poirier2010positron}
F~Poirier, I~Chaikovska, O~Dadoun, P~Lepercq, R~Roux, A~Variola, R~Chehab,
  R~Boni, S~Guiducci, M~Preger, et~al.
\newblock Positron production and capture based on low energy electrons for
  superb.
\newblock In {\em 1st International Particle Accelerator Conference (IPAC
  2010)}, pages 1650--1652. Joint Accelerator Conferences Website, 2010.

\bibitem{Liu:2021iay}
Jing-Dong Liu et~al.
\newblock {System design and measurements of flux concentrator and its
  solid-state modulator for CEPC positron source}.
\newblock {\em Nucl. Sci. Tech.}, 32(7):77, 2021.

\bibitem{lapik2004some}
R~Lapik, A~Yakutin, and P~Martyshkin.
\newblock Some problems of high field pulsed magnets development.
\newblock Technical Report BUDKER-INP-2004-73, 2004.

\bibitem{comsol}
Comsol multiphysics.
\newblock https://www.comsol.com.

\bibitem{PhysRevSTAB.8.072001}
A.~Moretti, Z.~Qian, J.~Norem, Y.~Torun, D.~Li, and M.~Zisman.
\newblock Effects of high solenoidal magnetic fields on rf accelerating
  cavities.
\newblock {\em Phys. Rev. ST Accel. Beams}, 8:072001, Jul 2005.

\bibitem{PhysRevAccelBeams.23.072001}
D.~Bowring, A.~Bross, P.~Lane, M.~Leonova, et~al.
\newblock Operation of normal-conducting rf cavities in multi-tesla magnetic
  fields for muon ionization cooling: A feasibility demonstration.
\newblock {\em Phys. Rev. Accel. Beams}, 23:072001, Jul 2020.

\bibitem{formela2021designing}
M~Formela, N~Hamann, K~Fl{\"o}ttmann, G~Moortgat-Pick, and S~Riemann.
\newblock Designing a plasma lens as a matching device for the ilc positron
  source.
\newblock {\em arXiv preprint arXiv:2105.14008}, 2021.

\bibitem{Paramonov2006}
V~Paramonov and K~Floettmann.
\newblock {Beam-Loading Effect in the Normal-Conducting {ILC} Positron Source
  Pre-Accelerator}.
\newblock In {\em Proceedings of LINAC 2006, Knoxville, Tennessee USA}, pages
  355--357, 2006.

\bibitem{Kuriki2019}
M~Kuriki, H~Nagoshi, T~Takahashi, K~Negishi, T~Okugi, T~Omori, M~Satoh,
  Y~Seimiya, J~Urakawa, and K~Yokoya.
\newblock {Beam loading compensation in the capture linac of {ILC} e-driven
  positron source}.
\newblock In {\em Proceedings of the 16th annual meeting of Particle
  Accelerator Society of Japan (PASJ2019)}, pages 776--779, 2019.

\bibitem{aune1979new}
B~Aune and RH~Miller.
\newblock New method for positron production at slac.
\newblock Technical Report SLAC-PUB 2393, Stanford Linear Accelerator Center,
  CA (USA), 1979.

\bibitem{HAN201983}
Y.L. Han, C.~Bayar, A.~Latina, S.~Doebert, D.~Schulte, and L.L. Ma.
\newblock Optimization of the {CLIC} positron source using a start-to-end
  simulation approach involving multiple simulation codes.
\newblock {\em Nuclear Instruments and Methods in Physics Research Section A:
  Accelerators, Spectrometers, Detectors and Associated Equipment}, 928:83--88,
  2019.

\bibitem{Zhao:2021tfi}
Yongke Zhao, Steffen D\"obert, Andrea Latina, and Lianliang Ma.
\newblock {A New Algorithm for Positron Source Parameter Optimisation}.
\newblock In {\em {12th International Particle Accelerator Conference~}}, 8
  2021.

\bibitem{ALLISON2016186}
J.~Allison, K.~Amako, J.~Apostolakis, et~al.
\newblock Recent developments in geant4.
\newblock {\em Nuclear Instruments and Methods in Physics Research Section A:
  Accelerators, Spectrometers, Detectors and Associated Equipment},
  835:186--225, 2016.

\bibitem{osti_877507}
A~Ferrari, P~R Sala, A~Fasso, and J~Ranft.
\newblock {FLUKA}: A multi-particle transport code.
\newblock In {\em CERN 2005-10, INFN/TC 05/11, SLAC-R-773}, 2005.

\bibitem{astra}
K.~Flöttmann.
\newblock Astra particle tracking code.
\newblock http://www.desy.de/~mpyflo/.

\bibitem{latina2016rf}
Andrea Latina et~al.
\newblock {RF-TRACK}: Beam tracking in field maps including space-charge
  effects. features and benchmarks.
\newblock In {\em Proceedings of the 28th Linear Accelerator
  Conference-LINAC16}, 2016.

\bibitem{gpt}
S.~B. van~der Geer and M.~J. de~Loos.
\newblock General particle trace.
\newblock www.pulsar.nl/gpt.

\bibitem{Deniau:ICAP12-THAAI3}
L.~Deniau.
\newblock Mad-x progress and future plans.
\newblock In {\em Proc. 11th Int. Computational Accelerator Physics Conf.
  (ICAP'12)}, pages 211--226. JACoW Publishing, Aug. 2012.

\bibitem{sad}
K.~Oide.
\newblock Sad.
\newblock http://acc-physics.kek.jp/SAD/index.html,
  https://github.com/KatsOide/SAD.

\bibitem{latina2012improvements}
A~Latina, E~Adli, B~Dalena, D~Schulte, and J~Snuverink.
\newblock Improvements in the placet tracking code.
\newblock In {\em Conf. Proc.}, volume 1205201, page MOPPC073, 2012.

\bibitem{Bacci:2016few}
Alberto Bacci, Vittoria Petrillo, and Marcello Rossetti~Conti.
\newblock {GIOTTO: A Genetic Code for Demanding Beam-dynamics Optimizations}.
\newblock In {\em {7th International Particle Accelerator Conference}}, pages
  3073--3076, 2016.

\bibitem{ansys}
Ansys.
\newblock https://www.ansys.com.

\bibitem{XAVIER1990278}
Xavier Artru.
\newblock A simulation code for channeling radiation by ultrarelativistic
  electrons or positrons.
\newblock {\em Nuclear Instruments and Methods in Physics Research Section B:
  Beam Interactions with Materials and Atoms}, 48(1):278--282, 1990.

\bibitem{PhysRevA.86.042903}
Vincenzo Guidi, Laura Bandiera, and Victor Tikhomirov.
\newblock Radiation generated by single and multiple volume reflection of
  ultrarelativistic electrons and positrons in bent crystals.
\newblock {\em Phys. Rev. A}, 86:042903, Oct 2012.

\bibitem{PhysRevAccelBeams.22.064601}
A.~I. Sytov, V.~V. Tikhomirov, and L.~Bandiera.
\newblock Simulation code for modeling of coherent effects of radiation
  generation in oriented crystals.
\newblock {\em Phys. Rev. Accel. Beams}, 22:064601, Jun 2019.

\bibitem{Baier_Katkov_Strakhovenko}
V.~N. Baier, V.~M. Katkov, and V.~M. Strakhovenko.
\newblock Radiation yield of high-energy electrons in thick crystals.
\newblock {\em physica status solidi (b)}, 133(2):583--592, 1986.

\bibitem{PhysRevLett.121.021603}
L.~Bandiera, V.~V. Tikhomirov, M.~Romagnoni, N.~Argiolas, E.~Bagli,
  G.~Ballerini, A.~Berra, C.~Brizzolari, R.~Camattari, D.~De~Salvador,
  V.~Haurylavets, V.~Mascagna, A.~Mazzolari, M.~Prest, M.~Soldani, A.~Sytov,
  and E.~Vallazza.
\newblock Strong reduction of the effective radiation length in an axially
  oriented scintillator crystal.
\newblock {\em Phys. Rev. Lett.}, 121:021603, Jul 2018.

\bibitem{newton2010rapid}
D~Newton et~al.
\newblock The rapid calculation of synchrotron radiation output from long
  undulator systems.
\newblock {\em Proceedings of IPAC2010, Kyoto, Japan}, 2010.

\bibitem{tanaka2001spectra}
Takashi Tanaka and Hideo Kitamura.
\newblock {SPECTRA}: a synchrotron radiation calculation code.
\newblock {\em Journal of synchrotron radiation}, 8(6):1221--1228, 2001.

\bibitem{Yokoya}
{K. Yokoya}.
\newblock {Users manual of {CAIN}, Version 2.42}.

\bibitem{PhysRevLett.121.044801}
Alexander Scheinker, Auralee Edelen, Dorian Bohler, Claudio Emma, and Alberto
  Lutman.
\newblock Demonstration of model-independent control of the longitudinal phase
  space of electron beams in the linac-coherent light source with femtosecond
  resolution.
\newblock {\em Phys. Rev. Lett.}, 121:044801, Jul 2018.

\bibitem{20.500.11850/385955}
Bayesian optimisation for fast and safe parameter tuning of {SwissFEL}.
\newblock In Winfried Decking, Harald Sinn, Gianluca Geloni, et~al., editors,
  {\em FEL2019, Proceedings of the 39th International Free-Electron Laser
  Conference}, pages 707 -- 710, s.l., 2019-11. JACoW Publishing.
\newblock 39th International Free Electron Laser Conference (FEL 2019);
  Conference Location: Hamburg, Germany; Conference Date: August 26-30, 2019;
  Conference lecture held on August 29, 2019.

\bibitem{PhysRevLett.126.104801}
S\"oren Jalas, Manuel Kirchen, Philipp Messner, Paul Winkler, Lars H\"ubner,
  Julian Dirkwinkel, Matthias Schnepp, Remi Lehe, and Andreas~R. Maier.
\newblock Bayesian optimization of a laser-plasma accelerator.
\newblock {\em Phys. Rev. Lett.}, 126:104801, Mar 2021.

\bibitem{chen1992coherent}
Pisin Chen and Robert~B Palmer.
\newblock Coherent pair creation as a positron source for linear colliders.
\newblock In {\em AIP Conference Proceedings}, volume 279, pages 888--895.
  American Institute of Physics, 1992.

\bibitem{PhysRevLett.79.1626}
D.~L. Burke, R.~C. Field, G.~Horton-Smith, et~al.
\newblock Positron production in multiphoton light-by-light scattering.
\newblock {\em Phys. Rev. Lett.}, 79:1626--1629, Sep 1997.

\bibitem{PhysRevLett.97.175003}
D.~K. Johnson, D.~Auerbach, I.~Blumenfeld, et~al.
\newblock Positron production by x rays emitted by betatron motion in a plasma
  wiggler.
\newblock {\em Phys. Rev. Lett.}, 97:175003, Oct 2006.

\end{thebibliography}






\end{document}